\documentclass[10pt,journal,twoside]{IEEEtran}
\usepackage{amsmath,amssymb,amsfonts,amsthm}
\usepackage{graphicx}
\usepackage{stfloats}
\usepackage{algorithm}
\usepackage{algpseudocode}
\usepackage{tabularx}
\usepackage{booktabs}
\usepackage{array}
\usepackage{multirow}
\usepackage{siunitx}
\usepackage[caption=false,font=footnotesize]{subfig}
\usepackage{enumitem}
\usepackage{url}
\title{TRACE: A Modular Framework for RIS-Assisted Channel Estimation and Differential Channel-Aware Reconfiguration}
\author{
Smriti Kumar,
Mansi Ambwani,
Arzad Alam Kherani,
Vimal Bhatia.
\thanks{S. Kumar and A. A. Kherani are with the Dept. of ECE, M. Ambwani is with Dept. of EE IIT Bhilai, India (e-mail: \{smritik, mansia, arzad.alam\}@iitbhilai.ac.in). V. Bhatia is with the Dept. of EE, IIT Indore, India (e-mail: vbhatia@iiti.ac.in).}
\thanks{This work was supported by the IIITB COMET Foundation under the Advanced Communication Systems vertical of the National Mission on Interdisciplinary Cyber-Physical Systems (NM-ICPS), Department of Science and Technology, Govt. of India.\\
This work is submitted to IEEE TCoM on 18/07/2026 for review.}
}
\begin{document}
\maketitle
\begin{abstract}
Reconfigurable Intelligent Surface (RIS) research tightly couples
channel-estimation, control and communication, yet existing
implementations often rely on fixed algorithmic pipelines,
making it difficult to compare alternative estimation, tracking
and communication strategies under identical conditions and to
study low-overhead RIS adaptation under time-varying channels.
This paper addresses both challenges through two complementary
contributions. First, it presents TRACE (Toolkit for RIS-assisted
channel-estimation, adaptive control and communication
experimentation), a modular socket-based framework that
decouples the transmitter, radio environment, controller and
receiver through separate control- and data-plane interfaces,
enabling reproducible evaluation across substitutable modules. Second, it proposes the differential
channel-aware RIS update (DCAR) algorithm, which estimates
channel perturbations from reduced probe observations using a
regularized differential update to reduce retraining overhead.
TRACE is validated through interchangeable minimum mean square
error and orthogonal matching pursuit channel-estimation,
Gaussian random walk and Gauss--Markov channel evolution,
BPSK, QPSK and 16-QAM modulation techniques and multiple RIS sizes,
without modifying the underlying framework. Within TRACE,
DCAR is observed to reduce pilot overhead and computational complexity while
maintaining beamforming performance close to Kalman-filter-based
tracking under time-varying channels.
\end{abstract}


\section{Introduction}
\begin{table}[!b]
\centering
\caption{Comparison of RIS channel-estimation methods and simulation frameworks}
\label{tab:sota_comparison}
\renewcommand{\arraystretch}{1.1}
\setlength{\tabcolsep}{1.5pt} 
\footnotesize
\begin{tabularx}{\columnwidth}{|c|>{\arraybackslash}X|c|c|c|c|c|c|c|}
\hline
\textbf{Work} & \textbf{CE / Tracking Method} & 
\rotatebox{90}{\textbf{AT}} & 
\rotatebox{90}{\textbf{POR}} & 
\rotatebox{90}{\textbf{MF}} & 
\rotatebox{90}{\textbf{CEM}} & 
\rotatebox{90}{\textbf{CDS}} & 
\rotatebox{90}{\textbf{ORR}} & 
\rotatebox{90}{\textbf{N.O.}} \\ \hline
\cite{refII}    & DFT-MMSE                                            & No      & No      & No      & No      & No      & No      & -- \\ \hline
\cite{refVII}   & Variational Bayesian CE                             & No      & Yes     & No      & No      & No      & No      & -- \\ \hline
\cite{refVIII}  & Low-Complexity LMMSE                                & No      & Yes     & No      & No      & No      & No      & -- \\ \hline
\cite{refIX}    & Structured Sparse OMP                               & No      & Yes     & No      & No      & No      & No      & -- \\ \hline
\cite{refX}     & SOMP + OG-SBL                                       & No      & Yes     & No      & No      & No      & No      & -- \\ \hline
\cite{refXI}    & Kalman-based Tracking                               & Yes     & Yes     & No      & No      & No      & Yes     & -- \\ \hline
\cite{refXXIII} & Adaptive Pilot Selection (CRLB/ Fisher-info) & Partial & Yes & No & No & No & Yes & -- \\ \hline
\cite{refXII}   & Adaptive Filtering                                  & Yes     & Partial & No      & No      & No      & Yes     & -- \\ \hline
\cite{refXIII}  & Partial RIS Probe                                 & Yes     & Yes     & No      & No      & No      & Yes     & -- \\ \hline
\cite{refXV}    & Adaptive CE for TV Channels                         & Yes     & Yes     & No      & No      & No      & Yes     & -- \\ \hline
\cite{refXVII}$\ddagger$   & Propagation Simulation                     & No      & No      & Partial & No      & No      & No      & \textbf{P} \\ \hline
\cite{refXVIII}$\star$ & Link-Level Simulation                       & Partial & No      & Yes     & Limited & No      & Custom  & \textbf{S} \\ \hline
\textbf{TRACE}$\dagger$ & \small DFT-MMSE, OMP, DCAR                 & Yes     & Yes     & Yes     & Yes     & Yes     & Yes     & \textbf{E} \\ \hline
\end{tabularx}

{\footnotesize~\raggedright
\textit{Keywords:}
AT: Adaptive Tracking;
POR: Pilot Overhead Reduction;
MF: Modular Framework;
CEM: Channel Evolution Modularity;
CDS: Control/Data Separation;
ORR: Online RIS Reconfiguration;
N. O.: Native Outputs (denotes the scope of native outputs generated by each work).\\
~\raggedright ~\textit{Note:} 1)$\ddagger$ SimRIS~\cite{refXVII} is a physical-channel/propagation simulator for RIS-assisted links. It natively generates propagation-level outputs (e.g., channel realizations, path loss and array responses), but does not provide integrated communication, channel-tracking or RIS-control functionality. 2)$\star$
Sionna~\cite{refXVIII} supports user-defined channel modeling through custom implementation; however, runtime-selectable channel-evolution modules and integrated channel-tracking functionality are not provided by default. 3)$\dagger$ TRACE's entries reflect software-level capability validated through socket-based simulation on a single host; they indicate architectural support for each criterion within the current implementation. 4) N.\ O.\ categories: \textbf{P} -- propagation-level outputs (e.g., channel realizations and path-loss data); \textbf{S} -- system/link-level outputs (e.g., BER/BLER); \textbf{E} -- experiment-level outputs (e.g., closed-loop tracking, controller statistics or pilot-overhead measurements); \textbf{--} -- not applicable, as the work proposes an estimation /tracking algorithm rather than a simulation framework.\par}
\end{table}
A reconfigurable intelligent surface is a surface made of many
small elements, each reflecting radio waves with an adjustable phase
shift to redirect signals without active transmitter or receiver
circuitry~\cite{refI}. By shaping these reflections, RIS can improve
coverage, reliability and energy efficiency, though this requires
channel-estimation, RIS phase control and data communication to
work together---functions that are usually built and tested
separately~\cite{refI}. RIS elements combine signals coherently at
the receiver only when the cascaded transmitter--RIS--receiver
(Tx--RIS--Rx) channel is accurately
estimated~\cite{refII,refIII}; since RIS has no RF chain, this
channel must instead be inferred from pilot transmissions and
Rx-side feedback~\cite{refII,refIV}.
However, training overhead scales with the number of RIS elements,
since orthogonal RIS probing configurations are sequentially applied
during CE~\cite{refII,refV,refVI}.

Existing RIS CE studies commonly use least-squares, ( minimum mean square error (MMSE) or
sparse-recovery methods with orthogonal pilots~\cite{refII,refIV,refV,refVII,refVIII,refIX,refX},
while time-varying channels are handled through Kalman filtering,
adaptive filtering or partial probing~\cite{refXI,refXXIII,refXII,refXIII,refXIV,refXV,refXVI}.
However, these methods are typically tested under a single channel model, so their behavior across different channel dynamics remains difficult to access and compare. They are also built and evaluated within algorithm-specific simulation setups, which makes it hard to compare different estimation, tracking and communication strategies fairly. A common framework with interchangeable
channel-estimation, modulation, channel-evolution and RIS-control
modules would allow such comparisons under identical operating
conditions.

Similarly, existing tools such as SimRIS~\cite{refXVII} and
Sionna~\cite{refXVIII} mainly model physical radio propagation
and static link performance. They cannot simulate continuous,
closed-loop interaction, where the transmitter, receiver and
RIS controller exchange real-time updates. Neither separates
control-plane signaling (probing commands, receiver feedback,
RIS phase updates) from data-plane traffic (pilot and payload
data), which limits the study of signaling overhead, retraining
policies and adaptive control-loop behavior independently of
the physical-layer waveform. 

TRACE addresses this gap through a common experimental framework with separate control- and data-plane interfaces, allowing CE, tracking and communication modules to be replaced without modifying the remaining pipeline. The proposed DCAR algorithm demonstrates the framework's tracking interface through a low-overhead differential update because coherent signal combining needs the RIS phase configuration to
stay aligned with the time-varying cascaded channel but as RIS size grows, frequent retraining for full channel re-estimation for slowly varying channels creates excessive pilot overhead. The main contributions of this work are:
\begin{itemize}
    \item \emph{The TRACE Framework:} A modular socket-based software architecture  with separate control- and data-plane interfaces
allows tracking strategies, propagation models and signaling
setups to be exchanged and evaluated under identical operating conditions.
  \item \emph{The DCAR Algorithm:} A regularized differential channel-aware RIS Update (DCAR) mechanism for adaptive RIS reconfiguration under time-varying channels; combines channel tracking with a regularized differential update, to suppress the noise amplification associated with partial channel observations and enables channel-state updates using reduced training resources. This enables low-overhead channel tracking while maintaining effective phase alignment accuracy under slowly varying channels.    
\item \emph{System-Level Validation:} TRACE pipeline
is evaluated across MMSE and orthogonal
matching pursuit (OMP) CE modules,
BPSK, QPSK and 16-QAM modulation schemes, Gaussian Random Walk and
Gauss--Markov channel-evolution models and RIS sizes
$N \in \{4, 16, 64, 256\}$, all under identical socket architecture and pilot
structures, showing the framework's ability to capture
estimator--environment interactions and algorithm tradeoffs
without modifying the underlying pipeline.
\end{itemize}
Table~\ref{tab:sota_comparison} summarizes recent RIS CE, adaptive tracking and simulation frameworks in relation to TRACE.
\section{TRACE Toolkit Framework}
\begin{figure*}[!htbp]
    \centering
\includegraphics[width=0.83\textwidth]{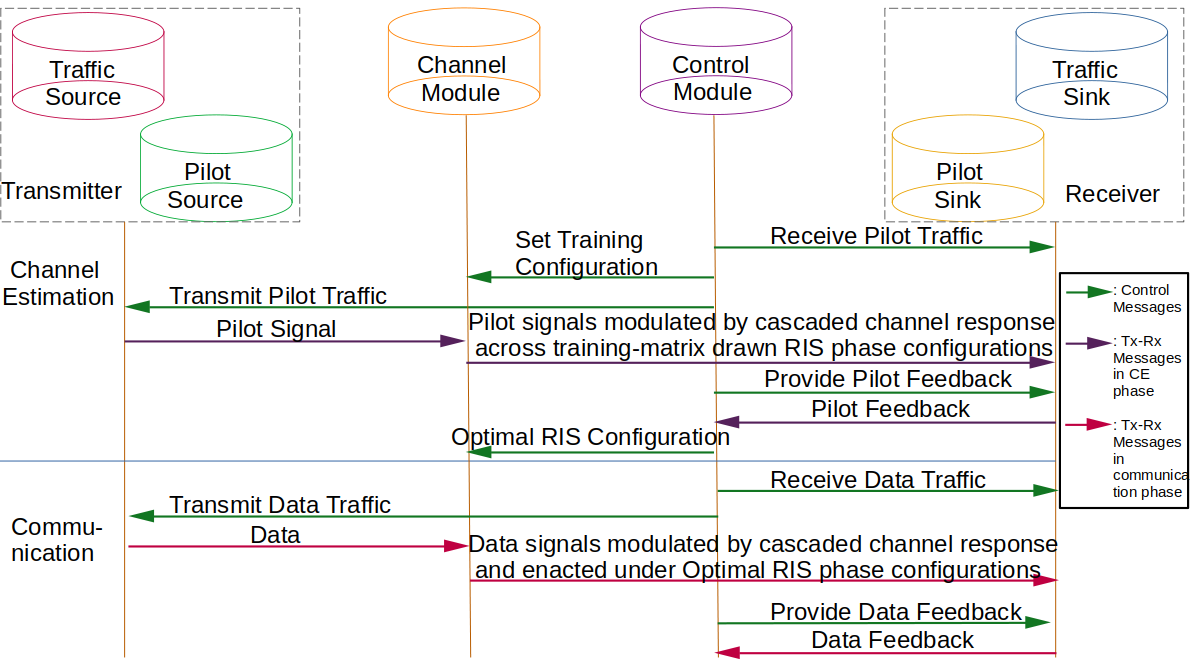}
    \caption{Socket-based control- and data-plane message exchange among the Tx, radio environment, controller and Rx modules.}
\label{fig:system_sequence}
    \end{figure*}
TRACE is a modular framework for RIS-assisted CE, communication 
and adaptive RIS reconfiguration organized into 
four logically independent modules: Tx, smart radio environment (also called 
channel module), controller and Rx, which interact through 
socket-based message exchange. Each message exchanged between 
modules is encapsulated in a serialized payload containing 
message-type headers and associated metadata. The current 
implementation realizes all four modules as software processes 
communicating over UDP sockets on a single host, serving as a software framework. Fig.~\ref{fig:system_sequence} shows the 
operational workflow and interaction among the framework 
components. UDP sockets provide a lightweight message interface between
independent modules, enabling explicit control- and data-plane
coordination while introducing only modest software overhead (see Sec.~\ref{discuss}).
\subsection{Operational Workflow}
TRACE operates through three sequential phases: channel
estimation, communication and channel tracking.

\subsubsection{Channel Estimation Phase} During CE, the RIS sequentially applies a
predefined set of probing configurations to acquire the
cascaded  Tx--RIS--Rx channel. For a RIS with \(N\) reflecting elements, the same number of pilot transmissions are used
during channel acquisition. Rx collects the corresponding pilot observations and forwards
them to the controller, which determines the cascaded channel estimate
and the optimal RIS phase configuration for coherent
signal combining.

\subsubsection{Communication Phase} Following CE, data transmission is performed
using the optimal RIS configuration. The
 Tx sends modulated symbols while the channel module maintains the selected RIS response. No additional
training overhead is incurred, as no pilot probing occurs during this phase.

\subsubsection{Tracking Phase} In slowly varying channels a full training phase at every interval introduces excess pilot overhead. To reduce this overhead, the tracking phase uses reduced channel observations to estimate only channel perturbations. Full retraining is triggered only when the accumulated channel drift 
exceeds a predefined threshold.

\subsection{Framework Architecture and Configurable Modules}
The four logically separated modules in TRACE interact through independent control- and data-plane interfaces. Control-plane signaling is used for RIS probing, Rx feedback and RIS configuration updates, whereas data-plane
communication carries pilot and payload observations, see Fig.~\ref{fig:Arch}.

The  Tx generates pilot and modulated symbols
corresponding to the selected communication scheme. The smart radio environment emulates RIS-assisted propagation through configurable channel and RIS responses. The Rx acquires pilot and payload observations and provides the feedback
required for CE and RIS control. The controller acts as the central coordinator by initiating RIS
probing, collecting Rx feedback, executing CE procedures, computing the RIS phase configurations and managing adaptive reconfiguration. 

TRACE allows independent configuration of CE, modulation and
RIS-control modules through module substitution and of
transmit power, pilot length, symbol count and RIS dimension
through a single shared configuration file edits
(configuration-level modifications). New estimation,
tracking or communication algorithms can be integrated by
replacing only the corresponding processing module while
preserving the existing socket message exchange with the
other modules.

\section{System Model and Channel Estimation}
This section presents the RIS-assisted signal model and the channel acquisition procedures used in TRACE. CE is performed via Discrete Fourier Transform(DFT)-based RIS probing followed by either MMSE estimation or OMP-based sparse recovery. This estimate is then used for RIS phase configuration and channel tracking.

\subsection{RIS-Assisted Channel Model}
\begin{figure}[!t]
    \centering
\includegraphics[width=0.82\linewidth]{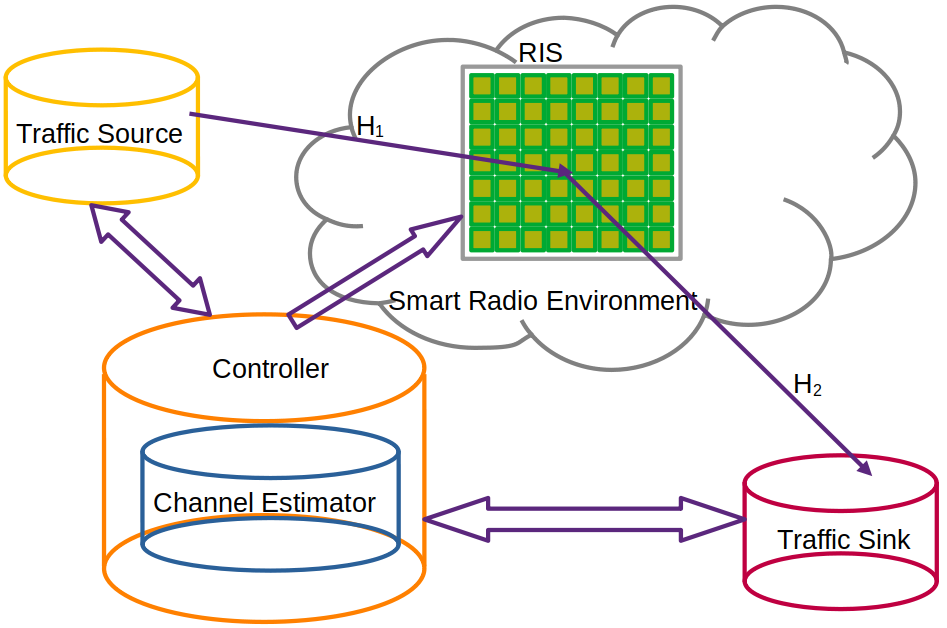}
   \caption{Tx--Rx link aided by an $N$-element RIS via channels $H_1$, $H_2$.}
    \label{fig:Arch}
\end{figure}
Consider the RIS-assisted communication system, see Fig.~\ref{fig:Arch}, consisting of a single-antenna  Tx and Rx aided by a RIS with $N$ passive reflecting elements. The  Tx to RIS channel is denoted by $H_1\!\in\!\mathcal{C}^{N \times 1}$, where $H_1\!=\![h_{11},\dots, h_{1N}]^{T}$ and the RIS-to-Rx channel is denoted by $H_2\!\in\!\mathcal{C}^{N \times 1}$, where $H_2^H\!=\![h_{21},\dots, h_{2N}]$ and there is no line-of-sight (NLoS) path, represented by the independent and
identically distributed (i.i.d.) channel coefficients modeled as
$h_{1i},h_{2i}\!\sim\!\mathcal{CN}(0,1)$~\cite{refII,refIII}. For static CE analysis, the  Tx--RIS
and RIS--Rx channels follow this
i.i.d. Rayleigh fading model. For dynamic channel-tracking
experiments, TRACE supports configurable
channel-evolution models, including Gaussian Random Walk (GRW)
and first-order Gauss--Markov (GM) processes (Sec.~\ref{sec:setup}).
If the $i^{{th}}$ RIS element applies an amplitude coefficient $\alpha_i\!\in\![0,1]$ and phase shift $\phi_i$ on the impinging signal, the $N$-element RIS reflection vector is expressed as $\theta\!=\![\alpha_1 e^{j\phi_1}, \dots, \alpha_N e^{j\phi_N}]^{T}.$  Within TRACE, the RIS is modeled as an ideal passive, phase-only reflecting surface, whose phase profile $\Phi\!=\!{diag}(\theta)$. 

Controller computes the RIS phase configuration for
coherent combining at receiver from the estimated cascaded channel,
assuming continuous phase control and no mutual coupling
between elements to isolate the estimation and tracking
algorithms from hardware non-idealities. For a transmitted symbol $x$, with transmit power $\mathcal{P}$ and $\omega\!\!\sim\!\mathcal{CN}(0,1)$ additive complex Gaussian noise, the received signal is given by
\begin{equation}
y=\sqrt{\mathcal{P}}H_2^H\Phi H_1 x+\omega,
\label{eq:1}
\end{equation}
where, $x$ can be a pilot or a data symbol. If $i^{{th}}$ RIS element's cascaded channel coefficient $g_i\!\!\triangleq\!\! h_{2i}h_{1i}$ and the cascaded channel vector $g\!\!=\!\![g_1, g_2, \dots, g_N]^{H}$, then \eqref{eq:1} expands to\footnote{The Hermitian transpose is used throughout for complex vectors: it correctly forms row-vector/bilinear-form representations~\eqref{eq:1} and for circularly-symmetric complex Gaussian vectors it is the only informative covariance, since the pseudo-covariance like $E[gg^T]=0$, see Sec.\ref{subsubsec:MMSE}~\cite{refII,refIV}.}
\begin{equation}
y = \sqrt{\mathcal{P}} \sum_{i=1}^{N} h_{2i}h_{1i}\alpha_i e^{j\phi_i} x + \omega 
  = \sqrt{\mathcal{P}} {g}^T {\theta} x + \omega.
\label{eq:rx_signal1}
\end{equation}
In the subsequent analysis, unit-average symbol energy,
unit-amplitude RIS reflection and normalized transmit power
are assumed without loss of generality, i.e., expectation
$E(|x|^2)=1$, $\alpha_i=1,\forall i$ and $P=1$.

\subsection{Channel Estimation Methods}
\label{sec:CE}
TRACE provides substitutable CE modules for cascaded RIS-assisted channel acquisition within a common communication and control framework. This implementation demonstrates two independently configurable CE methods: minimum mean square error estimation and orthogonal matching pursuit-based sparse recovery. Both integrate with the same DFT-based orthogonal probing
engine to estimate the cascaded channel required
for optimal RIS phase configuration. \footnote{To avoid ambiguity, the notation associated with RIS
configuration and probing matrices is distinguished as follows.
The symbol $\Phi={diag}(\theta)$ denotes the RIS phase
matrix applied during communication, where $\theta$ is the RIS
reflection vector. The matrix $\Phi_T$ denotes the full DFT
training matrix used during channel acquisition, while
$\Phi_{K_D}$ denotes the reduced DFT probing matrix employed
during DCAR updates. The symbols $\Phi_{11}$, $\Phi_{12}$,
$\Phi_{21}$ and $\Phi_{22}$ in Fig.~\ref{fig:dft_training}
represent illustrative DFT probing directions. Finally,
$\Phi^{{opt}}$ denotes the RIS phase configuration
computed from the estimated channel.}.

\subsubsection{DFT-Based RIS Training}
\label{subsec:DFT}
\begin{figure}[!t]
    \centering
\includegraphics[width=0.85\columnwidth]{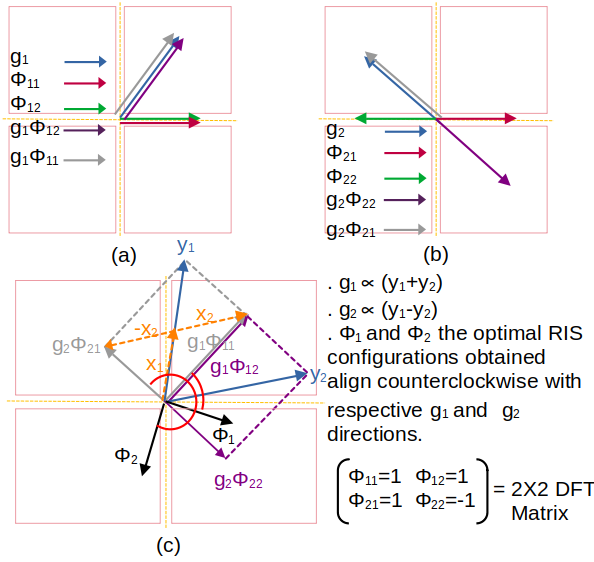}
    \caption{DFT-based RIS probing and spatial interpretation of cascaded
CE for a two-element RIS ($N=2$). (a) DFT operations for path~1, (b) DFT operations for path~2, (c) Geometric interpretation of the DFT-domain observations and
cascaded channel estimate reconstruction.\\
    \emph{Note:} $x_1$ and $x_2$ are DFT-domain observation
components obtained from projections of the received pilot
signal onto orthogonal DFT probing directions. The matrix entries omit the $1/\sqrt{N}$ factor of~\eqref{eq:dft_matrix} for clarity.}
\label{fig:dft_training}
\end{figure}
Channel acquisition is performed through sequential RIS probing using structured phase configurations. During the training phase\footnote{Although this work evaluates DFT-based probing, TRACE decouples the probing engine from the CE pipeline through a common observation interface. Alternative pilot/codebook designs can therefore be integrated without modifying the downstream estimation, phase-optimization, tracking or communication modules. DFT probing is adopted here to provide a controlled comparison between the interchangeable MMSE and OMP estimation modules.} orthogonal probing configurations are applied on RIS while  Tx sends pilot symbols. The received observations are then used to estimate the cascaded  Tx--RIS--Rx channel.

Several training strategies have been investigated for RIS-assisted CE, including canonical, Hadamard and DFT-based probing matrices \cite{refII,refIV,refVI}. Canonical probing activates one RIS element at a time, introducing high training overhead. Hadamard pilots 
are shorter and orthogonal and work better with MMSE at higher Signal--to--Noise--Ratio (SNR), but difficult to apply directly to RIS hardware, since controlling the amplitude is hard under unit-modulus limits~\cite{refII}. Hence, DFT-based pilots are adopted in TRACE because they satisfy the RIS phase-only constraint while preserving orthogonal probing~\cite{refVI}. The DFT training matrix \(\Phi_T\) is defined as:
\begin{equation}
[\Phi_T]_{i,j}
=
\frac{1}{\sqrt{N}}
\exp\left(
-j2\pi\frac{(i-1)(j-1)}{N}
\right), \Phi_T \in \mathcal{C}^{N \times N}
\label{eq:dft_matrix}
\end{equation}
where each column
$\phi_{T_j}=\Phi_T[:,j]$
represents one RIS probing configuration applied
sequentially during channel acquisition, requiring
$N$ pilot transmissions to probe all RIS dimensions.
Fig.~\ref{fig:dft_training} illustrates this process for
a two-element RIS ($N=2$), showing how orthogonal DFT
probing produces independent observations that are used
by the controller to reconstruct the cascaded
Tx--RIS--Rx channel estimate.

\subsubsection{MMSE Channel Estimation}
\label{subsubsec:MMSE}
The pilot observations collected during DFT probing are used
to estimate the cascaded channel. Let
$y_{rx}\in\mathcal{C}^{N\times1}$ denote the received pilot
observation vector and pilot symbol $x=1$, then the received signal is
\begin{equation}
y_{rx}=\mathbb{K}g+\omega,
\label{eq:training_model}
\end{equation}
where $g\!=\![g_1,\ldots,g_N]^H$ denotes the cascaded
 Tx--RIS--Rx channel vector,
$\mathbb{K}\!=\!\sqrt{\mathcal P}\Phi_Tx\!=\!\sqrt{\mathcal P}\Phi_T$ is the effective training matrix
determined by the RIS probing configurations,
and $\omega\!\sim\!\mathcal{CN}(0,\sigma^2I_N)$ denotes the stacked noise vector across the N probing configurations and $\sigma^2\!=\!1$ under the normalized noise assumption (thus the noise covariance matrix $\sigma^2 I_N\!=\!I_N$). Under the assumed i.i.d.\ Rayleigh fading model~\cite{refIV}, since each cascaded coefficient $g_i\!=\!h_{2i}h_{1i}$, where $h_{1i},h_{2i}\!\sim\!\mathcal{CN}(0,1)$ and
$h_{1i}\perp\!\!\!\perp h_{2i}$, yielding
$E[gg^H]\!=\!R_{gg}\!=\!I_N$, where $R_{gg}$ is the channel covariance matrix, following the covariance-based LMMSE formulation in~\cite{refII}. Hence, from \eqref{eq:training_model}, a linear estimator $\hat{g}_{{est}}\!=\!y_{rx}A$ is considered, where the weight matrix $A$ minimizes the
mean-squared estimation error (MSE)~\cite{refII, refIV}, $
e\!=\!E\!\left[
\left\|
g-y_{rx}A
\right\|_F^2
\right],
\label{eq:A_opt}$ where, $\left\| . \right\|_F$ denotes the Frobenius norm, which for a vector reduces to the Euclidean norm. The MSE in trace ($\operatorname{tr}(\cdot)$) form is
$
e\!=\!\operatorname{tr}\!\big(E[gg^H]\big)\!-\!\operatorname{tr}\!\big(E[gg^H] \mathbb{K}A\big)
    \nonumber\!-\!\operatorname{tr}\!\big(A^H \mathbb{K}^H E[gg^H]\big)   \nonumber\!+\!\operatorname{tr}\!\Big(
      A^H
      \big(
      \mathbb{K}^H E[gg^H]\mathbb{K}
      +\sigma^2I_N
      \big)
      A
      \Big).$
Setting $\partial e/\partial A\!=\!0$ to obtain $A = \underset{A}{\arg\min}\;{E}\left[ \left\| g\!-\!y_{{rx}} A \right\|_{F}^2\!\right]$:
\begin{align}
    A = \sqrt{\mathcal{P}} E[gg^H] \Phi_T^H \left( \mathcal{P} \Phi_T E[gg^H] \Phi_T^H + \sigma^2 I_N \right)^{-1}
    \label{eq:mmse_matrix}
\end{align}
substituting $E[gg^H]\!=\!I_N$, $\sigma^2\!=\!1$ and expressing the result in terms of the 
unit-modulus DFT matrix $F_T=\sqrt{N}\Phi_T$ physically applied by the RIS, the standard
closed-form MMSE channel estimate for $N$-pilot DFT-based training~\cite{refII} is

\begin{equation}
\hat{g}_{{est}}^H
=
\frac{
y_{rx}^H F_T \sqrt{\mathcal P}
}
{
N\mathcal P+1
}.
\label{eq:g_mmse}
\end{equation}
The result $\hat g_{{est}}\in C^{1 \times N}$ represents
the channel estimate and is forwarded to the RIS phase
optimization and channel-tracking modules, for computing the optimal RIS phase configuration
$\Phi^{opt}$ to maximize coherent signal combining
at the Rx.

\subsubsection{OMP-Based Sparse Channel Estimation}
\label{subsubsec:OMP}
\begin{figure}[!t]
    \centering
\includegraphics[width=\columnwidth]{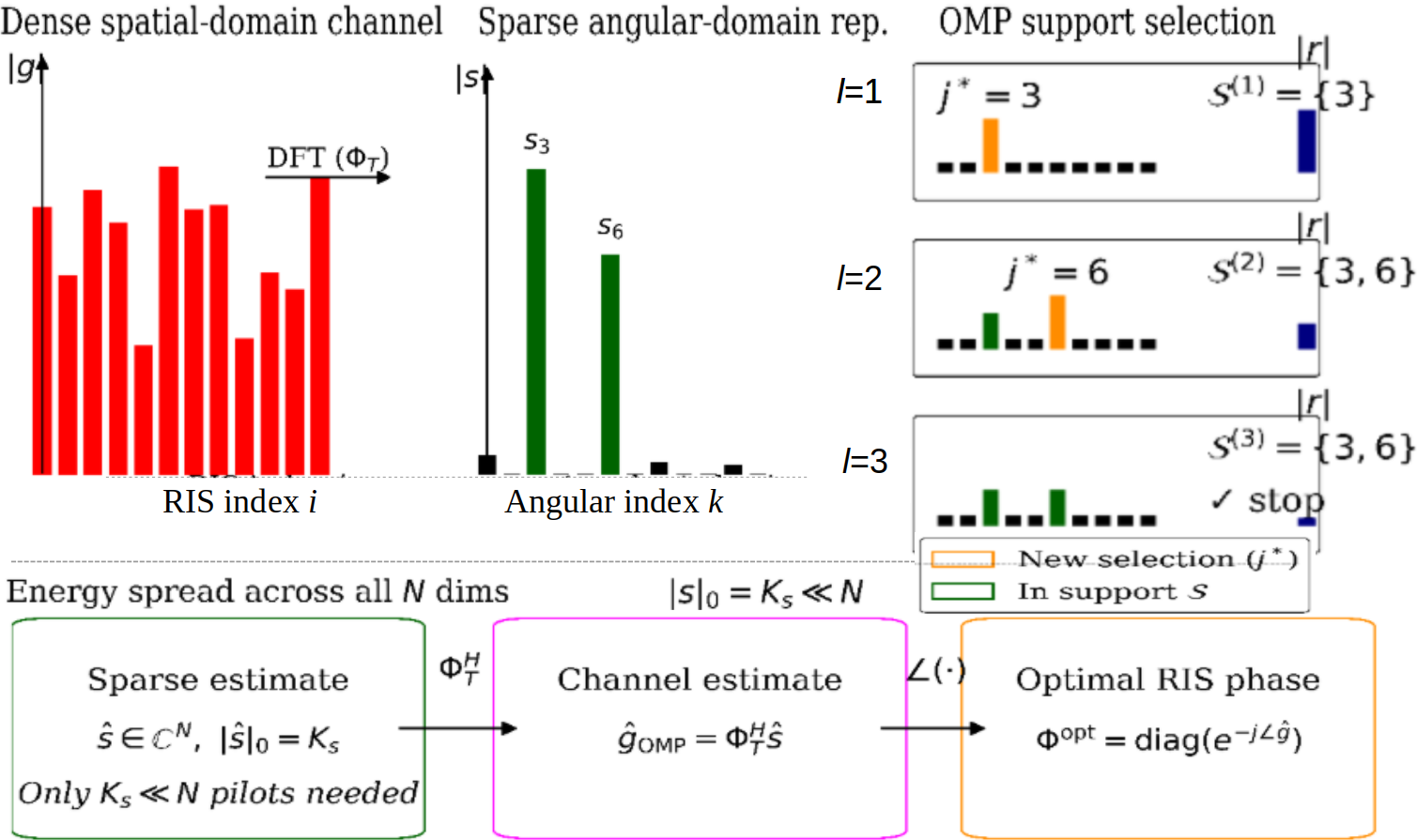}
    \caption{OMP-based sparse RIS channel estimation.}
    \label{fig:omp}
    \end{figure}
MMSE achieves optimal estimation when the channel covariance
matrix is known or estimated. In sparse propagation
environments like millimeter-wave and sub-THz channels,
however, only $K_s \ll N$ dominant angular paths are typically
present and accurate knowledge of the full covariance matrix
is generally unavailable~\cite{refIII}. OMP-based
sparse recovery instead exploits this structural sparsity directly,
without requiring channel covariance information.
It is implemented here  to demonstrate that TRACE supports interchangeable CE modules within a common
processing pipeline.

The pilot observation model is,
$y_{rx} = \sqrt{\mathcal{P}} \Phi_T g + \omega,$
where $\Phi_T\in\mathcal{C}^{N\times N}$ is the DFT probing matrix.
Dividing by $\sqrt{\mathcal{P}}$ gives the normalized observation $\tilde{y} \triangleq y_{rx}/\sqrt{\mathcal{P}}$:
\begin{equation}
\tilde{y} = \Phi_T g + \frac{\omega}{\sqrt{\mathcal{P}}}.
\label{eq:omp_norm}
\end{equation}
Since $\Phi_T$ is unitary, its columns form an orthonormal
basis for the DFT domain. The cascaded channel can therefore
be represented as
$g=\Phi_T^Hs$, where
$s\in\mathbb{C}^{N}$ denotes the DFT-domain coefficient vector.
For sparse propagation environments,
$s$ contains only $K_s\ll N$ significant coefficients.
Using $\varpi=\Phi_T$ as the sensing matrix,
\eqref{eq:omp_norm} becomes
\begin{equation}
\tilde{y} = \varpi s + \frac{\omega}{\sqrt{\mathcal{P}}}.
\label{eq:CS_model}
\end{equation}
OMP iteratively estimates the support and values of the $K_s$ significant
coefficients in $s$ as shown in 
Fig.~\ref{fig:omp}.

\paragraph{OMP Algorithm}
Starting from the initial residual
$r^{(0)}=\tilde{y}$ and the empty support set
$S^{(0)}=\emptyset$, OMP iteratively identifies the dominant
support and estimates the corresponding coefficients.
At each iteration $l=1,\ldots,K_s$:
\begin{enumerate}
\item \textbf{Correlation Vector and \emph{Matching} Step:}
Correlate the residual with each column of $\varpi$,
\begin{equation}
c^{(l)} = \varpi^H r^{(l-1)},
\label{eq:omp_corr}
\end{equation}
and select the index with the largest magnitude,
$j^*\!\!=\!\!\arg\max_j\! |c_j^{(l)}|.$
\item \textbf{Support and \emph{Pursuit} Update:}
\begin{equation}
S^{(l)} = S^{(l-1)} \cup \{j^*\}.
\label{eq:omp_support}
\end{equation}
\item \textbf{Least-Squares and \emph{Orthogonal} Projection:}
Re-estimate the non-zero coefficients over the current support,
\begin{equation}
\hat{s}_S^{(l)}
= \arg\min_{s_S}\|\tilde{y}-\varpi_S s_S\|^2,
\label{eq:omp_ls}
\end{equation}
where $\varpi_S$ contains the columns of $\varpi$
indexed by $S^{(l)}$ and update the residual,
\begin{equation}
r^{(l)} = \tilde{y} - \varpi_S\hat{s}_S^{(l)}.
\label{eq:omp_res}
\end{equation}
The residual lies in the orthogonal complement of
$\mathrm{span}(\varpi_S)$, so
$\varpi_S^H r^{(l)}=0$ holds exactly and residual
energy decreases monotonically with each iteration.
\item \textbf{Normalized Residual-Energy Thresholding:}
Terminate if $l=K_s$ or if residual energy is below the
normalized residual threshold $\eta$
\begin{equation}
\frac{\|r^{(l)}\|^2}{\|\tilde{y}\|^2} < \eta.
\label{eq:omp_stop}
\end{equation}
\end{enumerate}
After convergence, the full sparse vector $\hat{s}$ is formed
by placing $\hat{s}_S^{(K_s)}$ at the recovered support indices
and zero elsewhere.
The spatial-domain channel estimate is then
\begin{equation}
\hat{g}_{\mathrm{OMP}} = \Phi_T^H\hat{s}.
\label{eq:omp_g}
\end{equation}
$\hat{g}_{\mathrm{OMP}}$ is forwarded to the phase-optimization
and tracking modules through the same controller interface as
the MMSE estimate. In all reported experiments, OMP
terminates at $l\!\!=\!\!K_s$ iterations. The normalized residual
threshold $\eta$ in~\eqref{eq:omp_stop} is set to
$10^{-6}$ as a numerical safeguard because, under the
evaluated i.i.d.\ Rayleigh fading channel, the DFT-domain
representation is dense and prevents early residual
convergence.

\paragraph{Tradeoffs}
MMSE performs robustly under dense Rayleigh fading and at low
SNR by exploiting the full channel covariance, at $\mathcal{O}(N^2)$
complexity; OMP requires no covariance knowledge and scales as
$\mathcal{O}(NK_s)$, making it preferable when the channel is
genuinely sparse in the angular domain. Under the i.i.d.\ Rayleigh
model used here, however, the channel is dense in the DFT domain,
so OMP incurs a structural mismatch that grows with $N$ (evaluated
in Sec.~\ref{sec:mod_ce}). Both modules share the same DFT pilot
observations and controller interface and this interface --
including the DFT probing scheme itself (Sec.~\ref{subsec:DFT}) --
is substitutable without modifying the surrounding TRACE pipeline.
\subsection{Differential Channel-Aware RIS Update (DCAR)}
\label{subsec:dcar}
Maintaining RIS phase alignment under a slowly time-varying channel via repeated full-channel updates incurs significant, avoidable pilot overhead. Instead of reconstructing the full cascaded channel matrix at each interval, the DCAR 
algorithm tracks low-dimensional channel perturbations relative to a stored anchor estimate. Let 
$\hat{g}^{(t-1)} \in \mathcal{C}^{N \times 1}$ be the cascaded channel estimate from interval $t-1$. Under slow fading, the channel evolves between consecutive
blocks as
\begin{equation}
    g^{(t)} = g^{(t-1)} + \delta_g^{(t)},
    \label{eq:dcar_channel_evolution}
\end{equation}
where $\delta_g^{(t)}\!\in\!\mathcal{C}^{N\times1}$ represents the perturbation vector satisfying $\|\delta_g^{(t)}\|\!\!\ll\!\!\|g^{(t-1)}\|$ (verified in Sec.~\ref{sec:dcar_heatmap}). The index $t$ denotes successive channel-update intervals by DCAR probing. Channel is assumed constant within each interval and evolves only between adjacent intervals separated by block duration $T_s$.

Instead of executing a $N$-pilot sounding cycle, the controller triggers a compressed probe sweep using a
subset of $K_D\!\ll\!N$ orthogonal configurations selected from a sub-sampled DFT probing matrix 
$\Phi_{K_D}\!\in\!\mathcal{C}^{K_D\!\times\!N}$. The received probe observation vector 
$y_{K_D}^{{probe,(t)}}\!\!\in\!\mathcal{C}^{K_D\times1}$ during interval $t$ is expressed as
\begin{equation}
    y_{K_D}^{{probe,(t)}} = \sqrt{\mathcal{P}} \Phi_{K_D} g^{(t)} + \omega_{k_D},
    \label{eq:dcar_probe}
\end{equation}
where $\omega_{k_D}\!\sim\!\mathcal{CN}(0, \sigma_n^2 I_{K_D})$ is the additive white Gaussian noise vector
\footnote{$\sigma_n^2\!=\!\mathcal{P}/\gamma_0$ represents the probe noise variance. $\sigma_n^2$ is maintained as an 
explicit tracking parameter in the regularization framework for dynamic tuning across operating SNRs. $\gamma_0$ is the 
transmit-to-noise ratio; setting $\gamma_0\!=\!\mathcal{P}\!=\!1$ recovers the normalized case $\sigma^2\!=\!1$ used in Sec.~\ref{subsubsec:MMSE}.}. Following the Gauss--Markov channel evolution model, see Sec.~\ref{sec:setup},
the cascaded channel coefficient
$g_i^{(t)}\!\!=\!\!h_{2i}^{(t)}h_{1i}^{(t)}$
is approximated as a first-order AR(1) process (correlation coefficient $\rho$).
The per-element perturbation variance
$2(1\!\!-\!\!\rho)$ in this model (derived in Sec.~\ref{sec:dcar_heatmap},~\eqref{eq:delta_energy}), remains small in the high-correlation
regime ($\rho\!\approx\!1$) considered in DCAR. Using the historical anchor $\hat{g}^{(t-1)}$, the controller predicts the expected probe response 
$\tilde{y}_{K_D}^{{pred},(t-1)}\!\!=\!\!\sqrt{\mathcal{P}} \Phi_{K_D}\hat{g}^{(t-1)}$, 
subtracting which from the physical probe observation vector isolates the differential observation vector 
$\Delta y_{K_D}^{(t)}$:
\begin{equation}
    \Delta y_{K_D}^{(t)} = y_{K_D}^{{probe,(t)}} - \tilde{y}_{K_D}^{{pred},(t-1)} = \sqrt{\mathcal{P}} \Phi_{K_D}\delta_g^{(t)} + \omega_{k_D}.
    \label{eq:dcar_model}
\end{equation}
Unlike full retraining, DCAR exploits temporal correlation between
consecutive intervals by estimating only the channel perturbation
$\delta_g^{(t)}$ rather than reconstructing the entire cascaded channel.
Because \eqref{eq:dcar_model} forms an under-determined inverse problem
($K_D<N$), a direct back-projection amplifies measurement
noise.\footnote{Setting $\mu\!\!=\!\!1$ in \eqref{eq:dcar_recovery} yields the
unregularized back-projection with recovered noise covariance
$R_n\!\!=\!\!\left(\frac{N}{K_D}\right)^2
\frac{\sigma_n^2}{\mathcal P}\Phi_{K_D}^{H}\Phi_{K_D}$.
$\Phi_{K_D}$ has orthonormal rows, so
$\operatorname{tr}(\Phi_{K_D}^{H}\Phi_{K_D})\!\!=\!\!\operatorname{tr}(I_{K_D})\!\!=\!\!K_D$, giving
$\operatorname{tr}(R_n)\!\!=\!\!
\frac{N^2}{K_D}\frac{\sigma_n^2}{\mathcal P}$,
i.e., an increase $\frac{\operatorname{tr}(R_n)\big|_{K_D}}{\operatorname{tr}(R_n)\big|_{K_D=N}}\!\!=\!\!\frac{N^2/K_D}{N}\!\!=\!\!\frac{N}{K_D}$ in recovered noise energy, to compensate for only using a ${K_D}$-dimensional subspace.}
To avoid this noise inflation, the perturbation vector is recovered
using an unbiased back-projection estimator regularized by an
LMMSE-motivated scalar shrinkage factor. The perturbation covariance follows directly from the AR(1)
channel model in Sec.\ref{sec:setup}. Assuming uncorrelated spatial
perturbations (cf. Sec.~\ref{subsubsec:MMSE}) with per-element variance $2(1-\rho)$ gives
$R_{\delta\delta}=E[\delta_g^{(t)}(\delta_g^{(t)})^H]
=2(1-\rho)I_N$. The corresponding LMMSE estimator~\cite{refIV}
is
\begin{equation}
    \hat{\delta}_g^{(t)} = R_{\delta\delta} \Phi_{K_D}^H
    \left( \Phi_{K_D}R_{\delta\delta}\Phi_{K_D}^H
    + \sigma_n^2 I_{K_D} \right)^{-1}
    \Delta y_{K_D}^{(t)}.
    \label{eq:lmmse_full}
\end{equation}

Since $\Phi_{K_D}$ consists of $K_D$ rows of an $N\times N$
unitary DFT matrix, it satisfies
$\Phi_{K_D}\Phi_{K_D}^H\!=\!I_{K_D}$.
Under the assumed covariance
$R_{\delta\delta}\!=\!2(1\!-\!\rho)I_N$, the matrix inverse
in~\eqref{eq:lmmse_full} reduces to the standard scalar LMMSE
shrinkage under orthogonal probing
(cf.~\cite[Sec.~12.3]{refIXX}). The factor
$N/K_D$ compensates for the reduced observation energy when
$K_D\!\!<\!\!N$ DFT probing directions are used. The resulting
perturbation estimate is
\begin{equation}
    \hat{\delta}_g^{(t)}
    = \frac{\mu}{\sqrt{\mathcal{P}}}
    \frac{N}{K_D}
    \Phi_{K_D}^{H}
    \Delta y_{K_D}^{(t)},
    \label{eq:dcar_recovery}
\end{equation}
where
\begin{equation}
    \mu=\frac{K_D/N}{K_D/N+\lambda},
    \qquad
    \lambda=\frac{\sigma_n^2}{2(1-\rho)N}.
    \label{eq:dcar_alpha_lambda}
\end{equation}

The shrinkage factor $\mu$ controls the estimated perturbation contribution to the channel update. As
$\sigma_n^2\!\rightarrow\!0$, $\mu\!\rightarrow\!1$, recovering the
unregularized projection. As the noise level increases,
$\mu$ decreases, suppressing noise amplification caused by
partial probing\footnote{ $\mu$ varies
smoothly with the operating SNR, $\rho$, $N$ and $K_D$
as per~\eqref{eq:dcar_alpha_lambda}.}. Then the updated cascaded channel estimate is
\begin{equation}
    \hat{g}^{(t)}
    =
    \hat{g}^{(t-1)}
    +
    \hat{\delta}_g^{(t)},
    \label{eq:dcar_update}
\end{equation}
which is subsequently used to compute the RIS phase
configuration. To monitor tracking quality, the controller
computes the normalized innovation metric
\begin{equation}
    v^{(t)}
    =
    \frac{\|y_{K_D}^{\mathrm{probe},(t)}
    -\tilde{y}_{K_D}^{\mathrm{pred},(t-1)}\|^2}
    {\|\tilde{y}_{K_D}^{\mathrm{pred},(t-1)}\|^2}.
    \label{eq:dcar_metric}
\end{equation}
This monitoring metric evaluates the drift in the tracking loop. The theoretical characterization of this tracking error, alongside its validation using empirical distributions, is detailed in Sec.~\ref{sec:dcar_heatmap}. To prevent instantaneous noise variations from prematurely breaking the tracking loop, the metric is filtered using an 
exponential moving average (EMA):
\begin{equation}
    v_{{EMA}}^{(t)} = \beta v^{(t)} + (1-\beta)v_{{EMA}}^{(t-1)},
    \label{eq:dcar_ema}
\end{equation}
where $\beta\!\!\in\!\!(0,1)$ is the EMA smoothing coefficient. A full
retraining is triggered when either
$v_{EMA}^{(t)}\!\!>\!\!\tau_{DCAR}$, where $\tau_{DCAR}$ is the threshold or the maximum tracking
interval $T_{\max}$ is reached. Otherwise,
\eqref{eq:dcar_recovery}--\eqref{eq:dcar_update}
continue tracking the channel using differential updates. $\tau_{DCAR}$ serves as a safeguard against
abrupt channel changes that violate the differential tracking
assumption\footnote{$\tau_{DCAR}$ is robust to $\rho$
mis-specification too. If the true channel decorrelates
faster than assumed when computing $\lambda$ and $\mu$
in~\eqref{eq:dcar_alpha_lambda}, the increasing
probe residuals raise $v_{EMA}^{(t)}$, triggering a full
retrain discontinuing mismatched
regularization tracking.}. Under highly correlated fading,
$v_{EMA}^{(t)}$ remains below $\tau_{DCAR}$, making
$T_{\max}$ the primary retraining mechanism (Section~\ref{subsubsec:tau_role} examines this behavior).

DCAR operates independently of the underlying CE module.
A full channel estimate is obtained via MMSE or OMP at the
start of each deployment or after a retrain is triggered.
Using this estimate as an anchor, DCAR then tracks channel
perturbations round by round until a full retrain is triggered.

\section{Numerical Results}
\label{sec:results}
This section evaluates the performance of the TRACE framework, spanning pipeline validation, modularity demonstrations and 
characterization of the DCAR mechanism. All evaluations use the same socket-coordinated core architecture, where 
distinct operating modes are selected by independently varying substitutable modules and configuration-level parameters, without altering the surrounding socket architecture.

\subsection{Experimental Setup}
\label{sec:setup}
The execution framework is built around the four-module TRACE architecture, see (Fig.~\ref{fig:system_sequence}). The socket-
coordinated control plane manages CE, tracking and RIS reconfiguration, while the data plane routes 
pilot signals and payload transmissions across separate UDP ports. The software architecture remains unchanged across all evaluations, only configuration parameters and substitutable modules vary. The Tx--RIS ($H_1$) and RIS--Rx ($H_2$) links are modeled as
independent i.i.d. Rayleigh fading channels,
$H_1,H_2\sim\mathcal{CN}(0,I)$~\cite{refII,refIII},
yielding a normalized cascaded channel power
${E}[|g_i|^2]={E}[|h_{2i}|^2]{E}[|h_{1i}|^2]=1$.
Under this normalized channel model, the effects of large-scale
path loss and deployment geometry are absorbed into the baseline
transmit-to-noise ratio
$\gamma_0=\mathcal{P}/\sigma_n^2$,
consistent with the normalization commonly adopted in RIS simulation
studies~\cite{refII,refIII,refIV}. Consequently, varying $\gamma_0$
changes only the receiver noise variance $\sigma_n^2$, while
$H_1$, $H_2$, $N$ and the underlying channel statistics remain
unchanged.

The system operates under a discrete-time block-fading model where the channels, $H_1$ and $H_2$, remain constant within each periodic transmission block $t$—encompassing CE, RIS phase configuration and payload delivery—and evolve exclusively between successive blocks separated by the interval $T_s$~\cite{refII,refXI,refXV}. TRACE supports two temporal channel-evolution models:

$\bullet$ A non-stationary Gaussian Random Walk (GRW) model:
\begin{equation}
H_i^{(t)} = H_i^{(t-1)} + \nu_{grw} Z_i^{(t)},
\end{equation}
where $i\!\in\![1,2]$ and  $Z_i^{(t)}\!\!\sim\!\mathcal{CN}(0,I)$ serves as the innovation vector, representing the random channel variation introduced at interval $t$ and $\nu_{grw}$ scales this innovation step size. Due to its linearly evolving variance 
${Var}(H_i^{(t)})\!\!=\!\! {Var}(H_i^{(0)})+t\nu_{grw}^2$, the GRW model exhibits progressive distributional drift and 
serves as a non-stationary reference for tracking performance~\cite{refXX}.

$\bullet$ A stationary first-order Gauss-Markov (GM) autoregressive (AR(1)) model \cite{refXXI,refXIII,refXV}:
\begin{equation}
H_i^{(t)} = \rho H_i^{(t-1)} + \sqrt{1-\rho^2} Z_i^{(t)},
\label{eq:GM}
\end{equation}
where $i\!\!\in\!\![1,2]$ and the channel evolves as a temporally correlated time series governed by the coefficient $\rho$. Here, the independent 
innovation vector $Z_i^{(t)}\!\!\sim\!\mathcal{CN}(0,I)$ models the stochastic fading variations across successive blocks,
while the scaling factor $\sqrt{1\!-\!\rho^2}$ ensures the marginal distribution 
remains stationary over time. In this discrete-time representation, the temporal correlation coefficient is obtained from the classical Jakes model, $\rho = J_0(2\pi f_D T_s)$~\cite{refXXII}, as commonly adopted in RIS tracking studies~\cite{refXI,refXV}, where $J_0(\cdot)$ is the zeroth-order Bessel function of the first kind, $f_D=v_sf_c/c$ is the Doppler frequency, $v_s$ is the scatterer velocity, $f_c=2.45$ GHz denotes the carrier frequency\footnote{The selected 2.45 GHz carrier represents the widely used ISM band. The modular architecture of TRACE allows this parameter to be replaced too.} and $T_s=10$ ms. For the slow-fading scenarios considered in this work, the argument $2\pi f_DT_s$ remains below $2.4048$ the first zero of $J_0(\cdot)$, which keeps $\rho$ relevant to the slow-fading scenarios studied here\footnote{Using the small-argument approximation $J_0(x)\!\approx\!1-x^2/4$, the correlation becomes $\rho\!\approx\!1-\pi^2f_D^2T_s^2$, hence, increasing mobility $v_s$ or update interval $T_s$ reduces temporal correlation, consistent with wireless fading behavior.}. 

Regardless of the underlying channel-evolution model, the
incremental perturbation is defined as $\delta_g^{(t)} = g^{(t)} - g^{(t-1)}$, across channel 
configurations. Under the AR(1) approximation for $g^{(t)}$~\eqref{eq:dcar_channel_evolution}, this perturbation can be approximated as:
\begin{equation}
\delta g^{(t)} \approx (\rho - 1)g^{(t-1)} + \sqrt{1-\rho^2} z^{(t)}.
\label{eq:pert}
\end{equation}
As $\rho \rightarrow 1$, both $(\rho-1)$ and
$\sqrt{1-\rho^2}$ approach zero, so the perturbation energy
remains small relative to the channel energy,
$\|\delta_g^{(t)}\| \ll \|g^{(t-1)}\|$.
This supports the differential-update assumption used by DCAR
(Sec.~\ref{sec:mod_ce}).

To evaluate tracking under distinct conditions, the framework
is operated using three mobility profiles, with Doppler
frequencies and correlation coefficients $\rho = J_0(2\pi f_D T_s)$~\cite{refXXII}
obtained from the Jakes model for the selected scatterer
velocities. The coherence time $T_c \approx 0.423/f_D$~\cite{refXXV}
is reported for each profile to confirm $T_s$ remains well
within the coherence interval:
1)\textbf{Slow Fading ($\rho\!\approx\!\!0.9975$):} Stationary indoor environments (micro-mobility setups) with a scatterer speed of $v_s\!=\!0.5$ m/s, $f_D\!\approx\!4.08$ Hz and $T_c\!\approx\!104$ ms. 2)\textbf{Moderate Fading ($\rho\!\approx\!0.9901$):} Normal indoor walking speeds of $v_s\!\!=\!1.0$ m/s, $f_D\!\!\approx\!\!8.17$ Hz and $T_c\!\approx\!\!52$ ms. 3)\textbf{Fast Fading ($\rho\!\approx\!0.676$):} Faster outdoor pedestrian (light vehicular transit) at $v_s\!\approx\!2.43$ m/s, $f_D\!\approx\!19.89$ Hz and $T_c\!\approx21$ ms.

Unless otherwise stated, baseline evaluations are performed
for RIS dimensions $N\!\!\in\!\!\{4,16,64,256\}$,
$\gamma_0\!\!\in\!\![-30,0]$ dB and use MMSE estimation (better matched to the dense Rayleigh channel model).
The DCAR evaluation uses the GM channel-evolution model,
consistent with its formulation for temporally correlated
channels, while GRW demonstrates TRACE's
support for interchangeable channel-evolution modules.

\subsubsection*{Beamforming Gain} 
The beamforming (BF) gain $G_{BF}$ is evaluated as a dimensionless power ratio relative to a single-element baseline ($N=1$) 
\cite{refII}. Because the channel coefficients are normalized such that $E[|g_i|^2] = 1$, a single ideal element yields 
an expected power of exactly $1$ (corresponding to unit channel variance, with no array BF magnification). Thus, $G_{BF}$ measures the signal power magnification over this single-element reference 
baseline and is defined as:
\begin{equation}
G_{BF} = \left|\sum_{i=1}^{N} \phi^{{opt}}_i  g_i \right|^2,
\label{eq:bf_gain_def}
\end{equation}
which is expressed in decibels as $G_{BF}{(dB)} = 10\log_{10}G_{BF}$. The phase configuration is set to 
$\phi^{{opt}}_i = e^{-j\angle\hat{g}_i}$, where $\hat{g}_i$ is the $i$-th element of the estimated channel vector. 
Under perfect channel state information, all phases align perfectly, reducing \eqref{eq:bf_gain_def} to the maximum 
possible gain $G_{BF,\max} = \left(\sum_{i=1}^{N}|g_i|\right)^2$. Beamforming gain is used as the primary tracking metric because it directly reflects the quality of RIS phase alignment and the resulting coherent-combining performance at the receiver. The effective received SNR (dB) after applying BF gain is:
\begin{equation}
\gamma_{\mathrm{eff}}(dB) = \gamma_0 + G_{BF}(dB),
\label{eq:effective_snr}
\end{equation}

For a large number of elements $N$, the expectation expands to 
$E\bigl[(\sum_{i=1}^N|g_i|)^2\bigr] = N^2(E[|g_i|])^2 + N {Var}(|g_i|)$. Because the $N^2$ term dominates as the 
array grows, the expected maximum gain satisfies $E[G_{BF,\max}] \approx N^2\bigl(E[|g_i|]\bigr)^2$. For the cascaded 
channel model($g_i=h_{2i}h_{1i}$, where $h_{1i},h_{2i}\sim\mathcal{CN}(0,1)$ and
$h_{1i}\perp\!\!\!\perp h_{2i}$) the magnitude is given 
by $|g_i| = |h_{1i}||h_{2i}|$. Due to this independence, the first two moments are explicitly evaluated as 
$E[|g_i|] = E[|h_{1i}|]E[|h_{2i}|] = \frac{\sqrt{\pi}}{2} \cdot \frac{\sqrt{\pi}}{2} = \frac{\pi}{4}$ and 
$E[|g_i|^2]=1$. Substituting these moments yields the expected maximum 
BF gain $E[G_{BF,\max}] \approx N^2\left(\frac{\pi}{4}\right)^2$.

\subsubsection*{Monte Carlo Methodology}
Performance metrics are reported as sample means over $M$ independent channel realizations. Statistical uncertainty is quantified using symmetric 95\% confidence intervals (CIs),
$\bar{x} \pm 1.96\,(\hat{\sigma}/\sqrt{M})$, where $\bar{x}$ and $\hat{\sigma}$ denote the sample mean and standard deviation, respectively. The representative tracking trajectories in Sec.~\ref{sec:validation} show a single simulation run rather than an ensemble average.

\subsubsection*{Default DCAR parameters}
By default, the DCAR mechanism utilizes $K_D = \lfloor N/4 \rfloor$ probe configurations, an EMA smoothing coefficient $\beta = 0.2$, a variation threshold $\tau_{{DCAR}} = 0.75$ and a forced-retrain interval $T_{\max} = 20$ rounds. 

\subsection{Framework Validation}
\label{sec:validation}
\begin{figure}[!htbp]
    \centering
    \subfloat[]{\includegraphics[width=0.811\columnwidth]{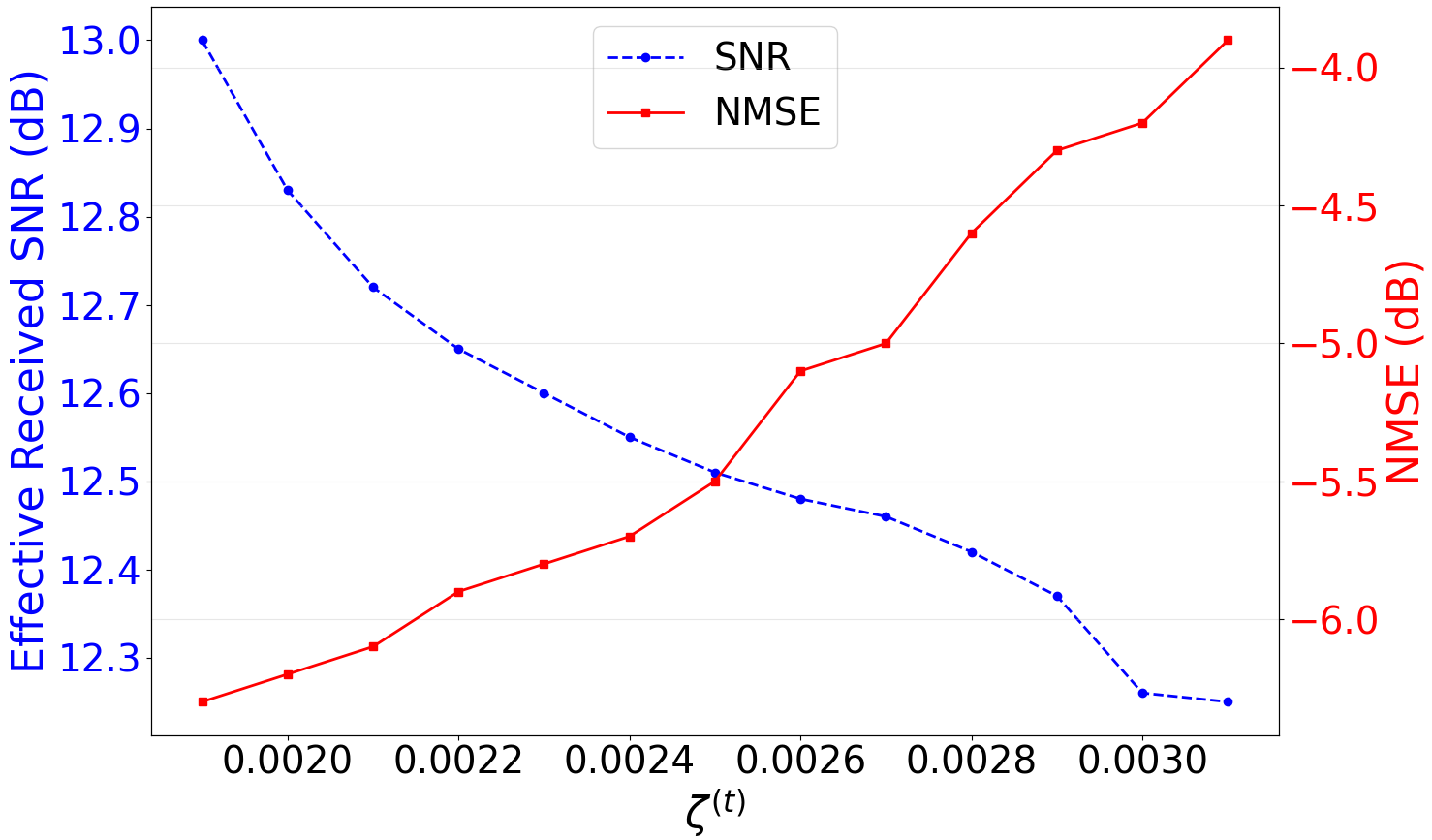}\label{fig:validation}}
    \\ \vspace{-0.5em}
    \subfloat[]{\includegraphics[width=0.811\columnwidth]{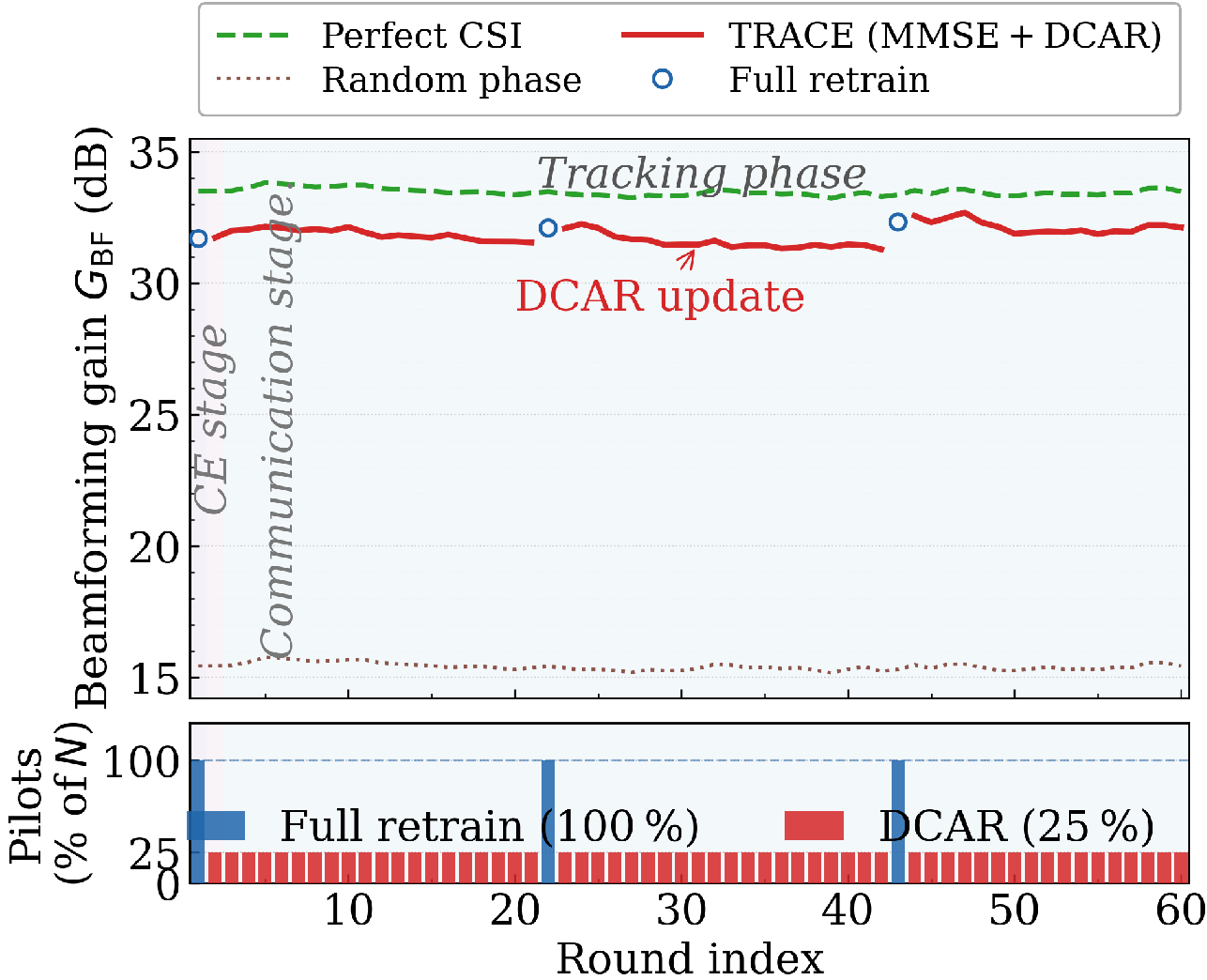}\label{fig:dcar_slow}}
    \\ \vspace{-0.5em}
    \subfloat[]{\includegraphics[width=0.811\columnwidth]{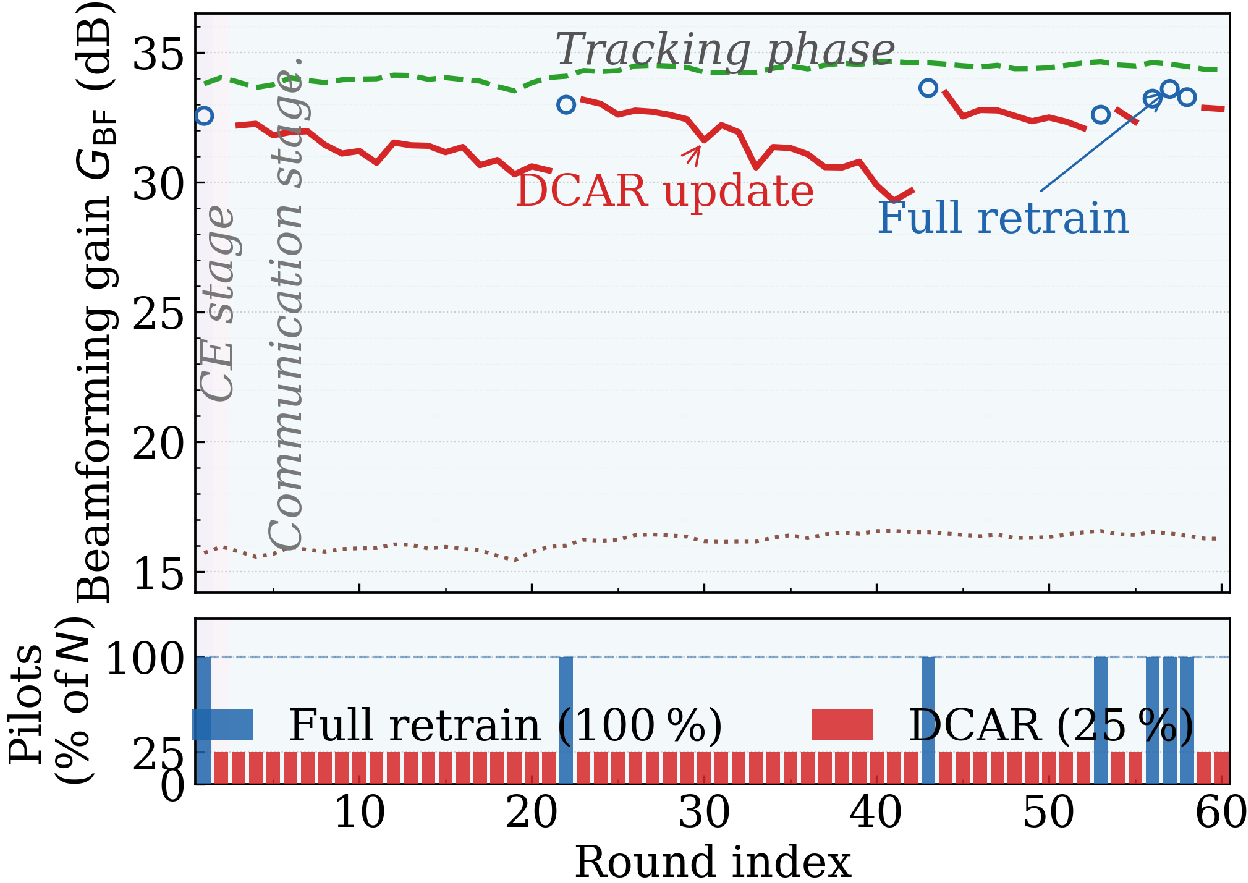}\label{fig:dcar_moderate}}
    \\ \vspace{-0.5em}
    \subfloat[]{\includegraphics[width=0.811\columnwidth]{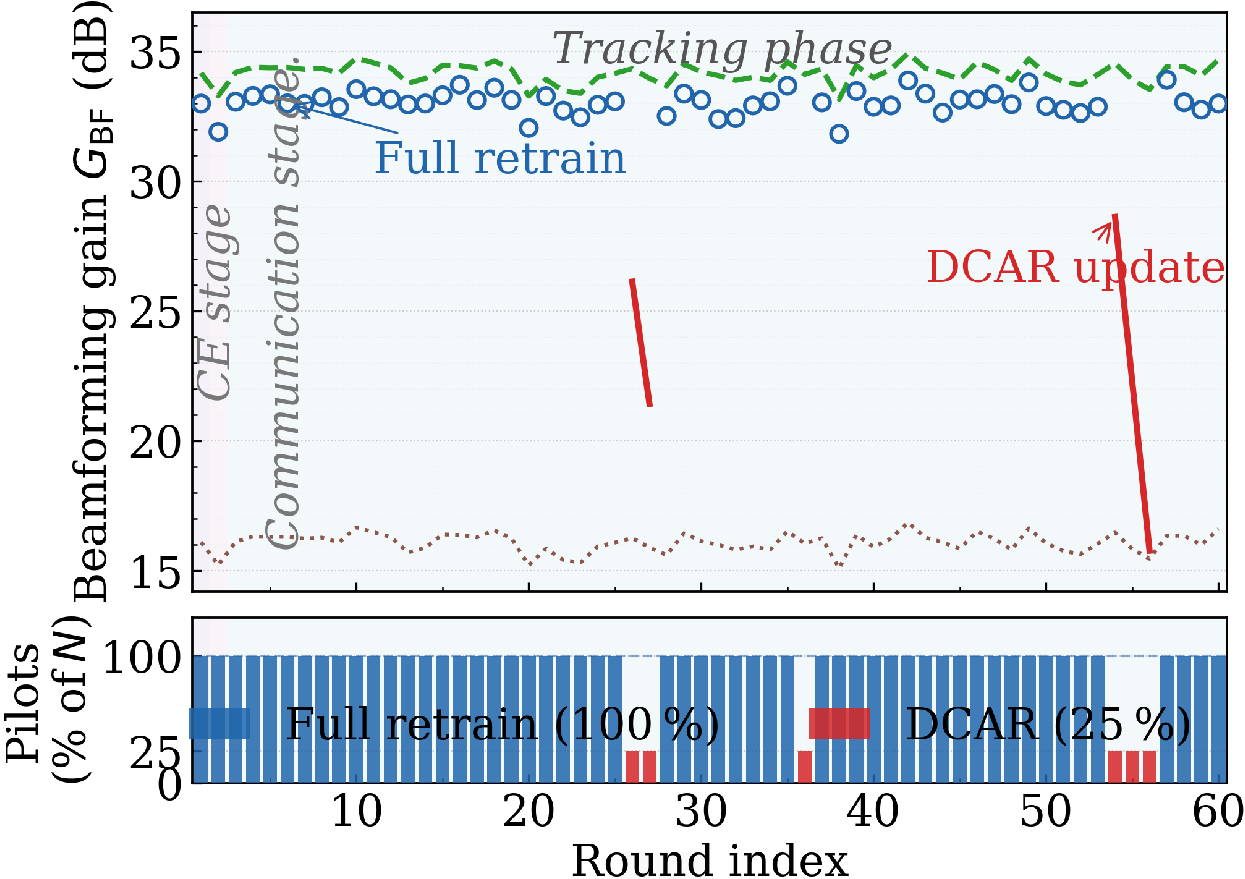}\label{fig:dcar_fast}}
\caption{Framework validation and tracking behavior under $N=64$, $\gamma_0=-20$ dB, $K_D=\lfloor N/4\rfloor$.
(a) Effective received SNR and NMSE versus $\zeta^{(t)}$ over 500 rounds under slow GM 
fading ($\rho\approx0.9975$). (b)--(d) Representative DCAR tracking 
trajectories and pilot overhead per round under slow 
($\rho\approx0.9975$), moderate ($\rho\approx0.9901$) and fast 
($\rho\approx0.676$) fading, respectively.}
\label{fig:dcar_tracking_examples}
\end{figure}
This experiment validates the system-level consistency of the TRACE pipeline by tracking how channel dynamics propagate through the estimation and BF stages; performance is evaluated using the relative channel perturbation:
\begin{equation}\zeta^{(t)} = \frac{|{\delta}_g^{(t)}|^2}{|{g}^{(t-1)}|^2}.\label{eq:validation_perturbation}\end{equation} 
While $\zeta(t)$ only provides a scalar summary of relative perturbation magnitude, it does not fully characterize the channel evolution and is used here  as an indicator of tracking loop reliability not as a complete channel descriptor.
Over $500$ rounds of
MMSE estimation under slow GM fading, $\zeta^{(t)}$, estimation
NMSE and effective received SNR were recorded each round and
sorted by increasing $\zeta^{(t)}$ to show how perturbations
propagate through the estimation and communication stages. As
Fig.~\ref{fig:dcar_tracking_examples}(a) shows, larger
perturbations coincide with higher NMSE and lower effective
SNR, confirming that channel evolution degrades estimation
accuracy and the BF gain.

Fig.~\ref{fig:dcar_tracking_examples}(b)--(d) show representative tracking trajectories under slow, moderate and fast GM
fading. In the slow-fading example ($\rho\!\approx\!0.9975$), DCAR maintains
BF gain within approximately $1$ dB of the full-retrain baseline for about
$15$--$18$ rounds before retraining, consistent with the differential update
assumption $\|\delta_g^{(t)}\|\!\ll\!\|g^{(t-1)}\|$. In the moderate-fading
example ($\rho\!\approx\!0.9901$), channel drift accumulates more rapidly,
leading to retraining after roughly $8$--$12$ rounds and a correspondingly
higher pilot fraction. The fast-fading example
($\rho\!\approx\!0.676$) shows rapid degradation, with DCAR reverting toward
full retraining at nearly every interval, reflecting that the algorithm is
primarily intended for slowly varying channels. Together, these representative
trajectories illustrate the expected relationship between channel correlation,
retraining frequency and pilot overhead across the evaluated fading
conditions.
\subsection{Modularity Demonstration}
\label{sec:modularity}
This subsection demonstrates TRACE's modularity through
independent substitution of CE modules
(MMSE and OMP), channel-evolution models (Gaussian Random
Walk and Gauss--Markov) and modulation schemes (BPSK, QPSK
and 16-QAM), while preserving the same socket architecture,
pilot structure and data pipeline. RIS size is varied
independently through configuration-level modifications.
\subsubsection{CE Module (MMSE vs.\ OMP)}
\label{sec:mod_ce}
To demonstrate TRACE's modularity under identical operating conditions, the CE core is interchanged between MMSE estimation (Sec.~\ref{subsubsec:MMSE}) and OMP sparse recovery (Sec.~\ref{subsubsec:OMP}). Both modules process the same DFT pilot observations and feed a common phase-optimization block. To ensure a consistent comparison, the OMP sparsity order and the DCAR probe count are bounded equally as $K_s = K_D = \lfloor N/4 \rfloor$. Fig.~\ref{fig:overall_results}(a) plots the mean BF gain against the RIS size $N$. The MMSE estimator closely tracks the theoretical bound, reaching $44.85$~dB ($\pm0.02$~dB, 95\% CI) at $N\!\!=\!\!256$, which is within $1.18$~dB of the Monte Carlo (MC) maximum $46.03$~dB. On the other hand, OMP achieves only $21.58$~dB at $N\!\!=\!\!256$, with the performance gap widening from $3.7$~dB at $N\!\!=\!\!4$ to $23.3$~dB at $N\!\!=\!\!256$.

This divergence stems from the mismatch between OMP's sparse
reconstruction model and the evaluated i.i.d.\ Rayleigh fading
channel. Under rich scattering, the channel is full-rank in the
angular domain. The unitary DFT probing matrix distributes pilot
energy uniformly while satisfying the RIS phase-only
constraint~\cite{refVI}, making it well suited to the rich-fading
MMSE framework. In contrast, OMP reconstructs only $K_s$
dominant paths, discarding significant multipath energy and
causing phase errors to increase with $N$. Since all other
modules remain unchanged, this divergence isolates the
estimator--channel mismatch itself, demonstrating TRACE's
ability to compare interchangeable estimation modules under
identical conditions; a controlled evaluation under sparse
channel models is left for future work~\cite{refXXIII,refXV}.
\subsubsection{Modulation Module}
\label{sec:mod_modulation}
To demonstrate modulation-level modularity, TRACE is
evaluated using BPSK, QPSK and 16-QAM under MMSE
estimation ($N=64$) and slow GM fading
($\rho\approx0.9975$). Fig.~\ref{fig:overall_results}(b)
plots the average BER against the effective received
SNR~\eqref{eq:effective_snr}. The measured BER closely follows the corresponding theoretical
trend; the residual gap is dominated by channel-estimation
error (Table~\ref{tab:cee_bins}), with additional low-SNR
deviation from effective-SNR-bin averaging and high-SNR
deviation from finite Monte Carlo sampling. BPSK
achieves the lowest BER, followed by QPSK and 16-QAM.
Since only the modulation module is changed while the
RIS configuration, CE and tracking pipeline remain
unchanged, the results show that different
modulation formats can be interchanged without
modifying the remaining framework.
\subsubsection{Channel Evolution Module}
\label{sec:mod_channel}
\begin{figure}[!t]
    \centering
    \subfloat[{}]{\includegraphics[width=0.85\columnwidth]{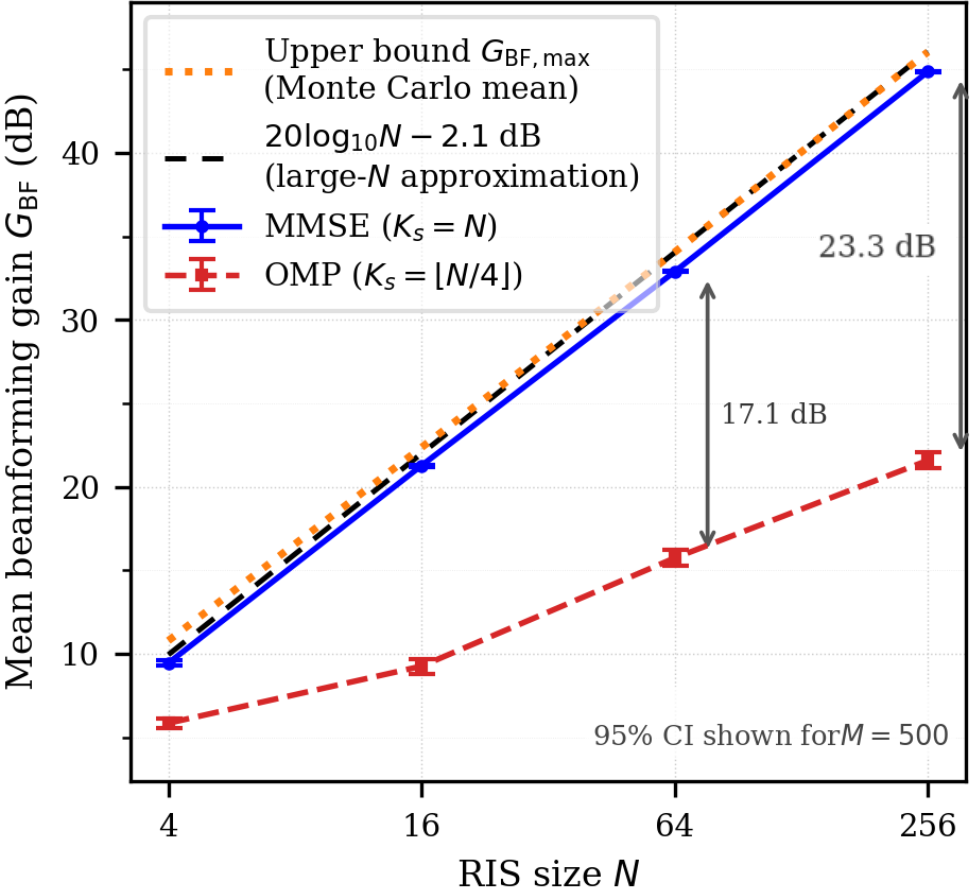}\label{fig:mod_ce}}%
    \\
    \vspace{-0.8em}
    \subfloat[{}]{\includegraphics[width=0.9\columnwidth]{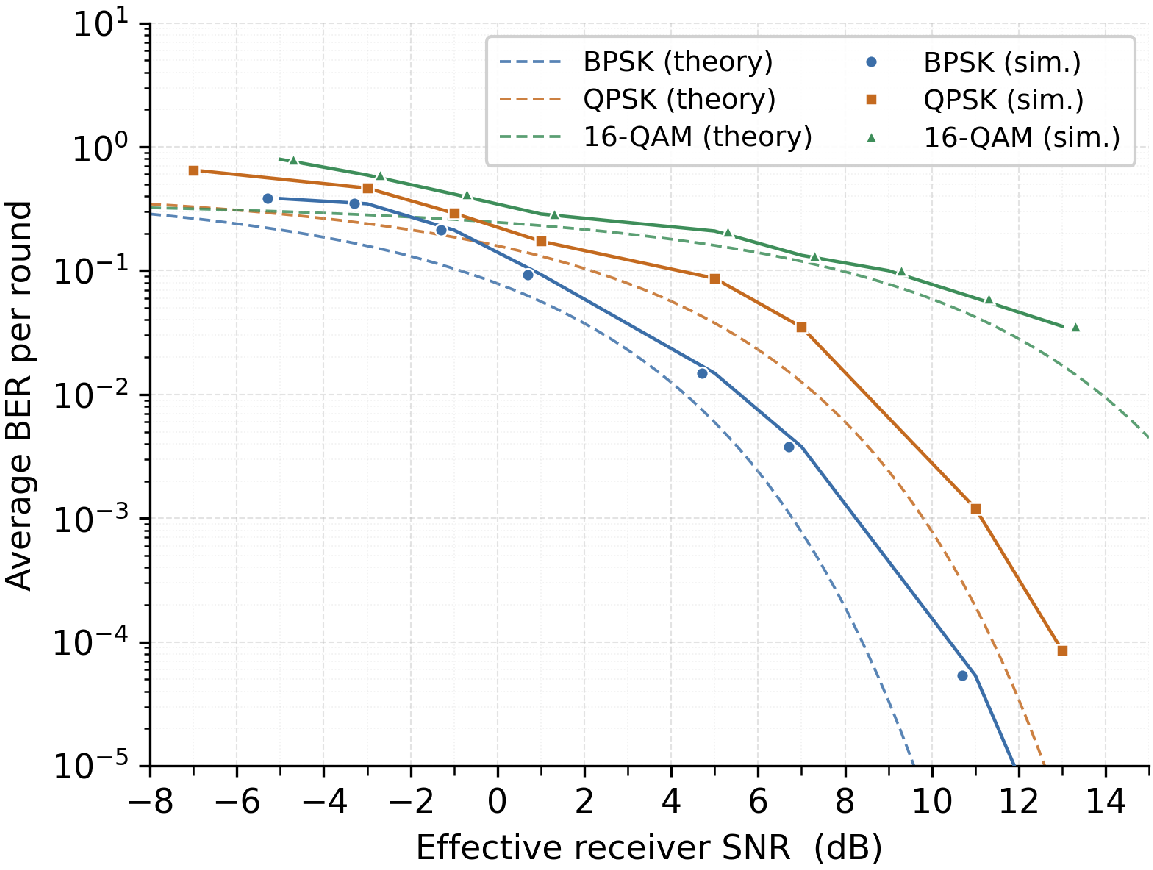}\label{fig:mod}}%
    \\
    \vspace{-0.8em}
    \subfloat[{}]{\includegraphics[width=0.9\columnwidth]{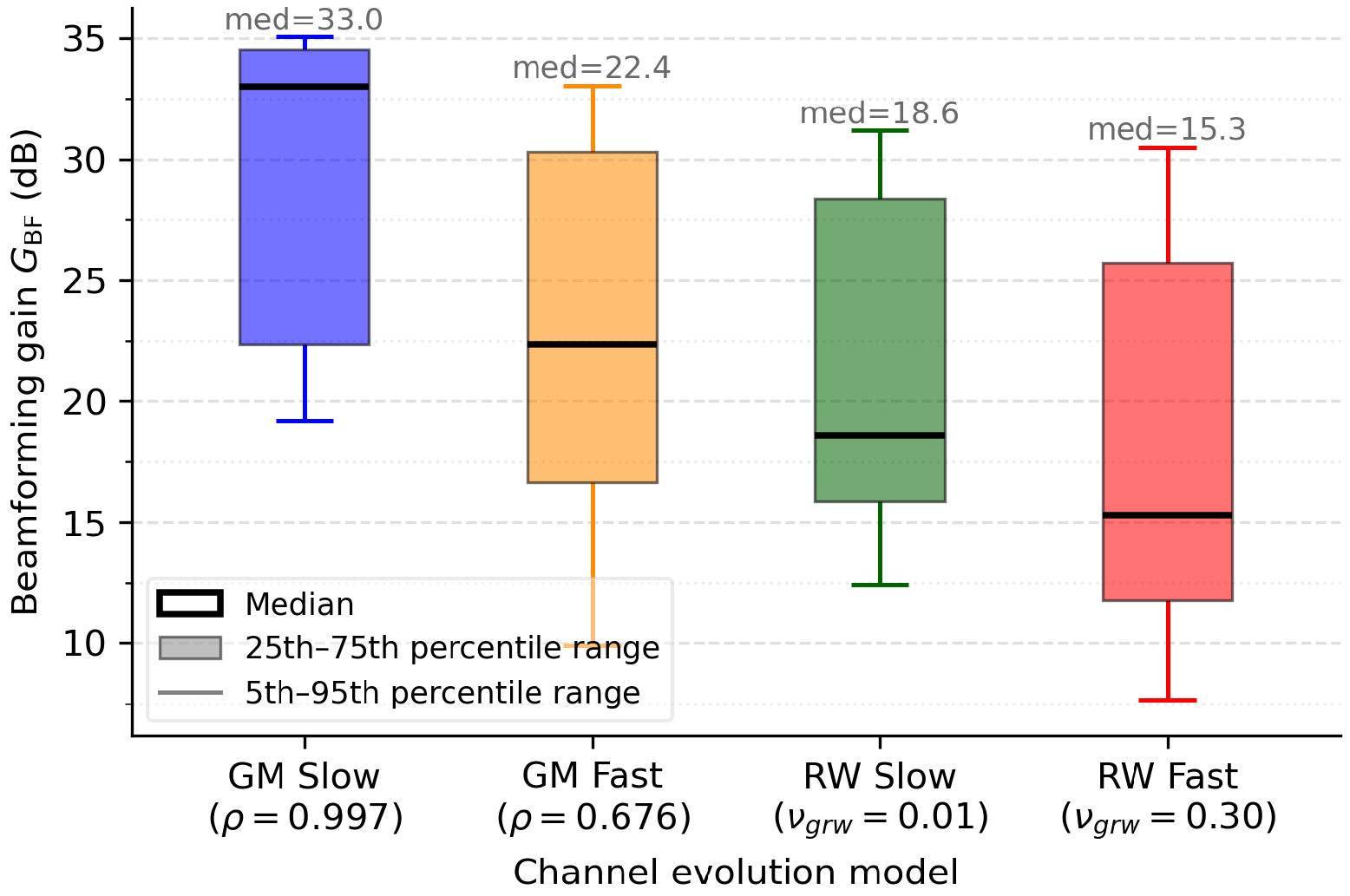}\label{fig:mod_channel}}%
    \caption{Modularity demonstration of TRACE framework: (a) Mean BF gain versus RIS size $N$ for MMSE and
OMP CE, QPSK modulation, $\gamma_0=-20$ dB and GM slow fading ($\rho\approx0.9975$, $v_s=0.5$ m/s); (b) Average per-round BER versus effective received SNR $\gamma_{\mathrm{eff}}(dB)$;(c) BF-gain distributions obtained under different
channel-evolution configurations.}
\label{fig:overall_results}
\end{figure}

Fig.~\ref{fig:overall_results}(c) compares the BF-gain 
distributions obtained under various channel-evolution models (Sec.~\ref{sec:setup}): 
GM slow ($\rho\!\!=\!\!0.9975$), GM fast ($\rho\!\!=\!\!0.676$), 
GRW with $\nu_{grw}\!\!=\!\!0.01$ and GRW with $\nu_{grw}\!\!=\!\!0.30$. 
Each scenario is evaluated across $500$ independent MC channel realizations,  for MMSE CE, $N\!\!=\!64$, $\gamma_0\!\!=\!-20$\ dB and QPSK modulation. The GM slow configuration achieves the highest median BF gain of $33.0$ dB. 
Under faster channel evolution GM fast the median gain drops
to $22.4$ dB, confirming that rapid temporal variations degrade the alignment of 
previously updated RIS states. The non-stationary GRW models yield even lower median gains of 
$18.6$ dB for $\nu_{grw}=0.01$ and $15.3$ dB for $\nu_{grw}=0.30$. 
It is observed that the broader percentile spreads under higher perturbations, indicating
the significant round-to-round signal fluctuations due to larger channel drift.

The observed BF-gain differences arise solely from the selected
channel-evolution model, since all other TRACE modules remain
unchanged. This demonstrates independent interchangeability of
the propagation model, enabling consistent comparison of
estimation, tracking and communication algorithms under
different channel conditions.
\begin{table}[!t]
\centering
\caption{Average channel-estimation error (CEE) per effective-SNR
bin for the modulation evaluation in Fig.~\ref{fig:overall_results}(b).}
\label{tab:cee_bins}
\renewcommand{\arraystretch}{1.1}
\setlength{\tabcolsep}{1pt}
\begin{tabularx}{\columnwidth}{|X|c|c|c|}
\hline
\textbf{Bin (Mean $\gamma_{\mathrm{eff}}$ (dB))} &
\textbf{BPSK CEE (dB)} &
\textbf{QPSK CEE (dB)} &
\textbf{16-QAM CEE (dB)} \\
\hline
$-5$ & --- & $-2.5$ & $-3.28$ \\
\hline
$-3$ & $-2.1$ & $-2.4$ & $-3.25$ \\
\hline
$-1$ & $-2.2$ & $-2.25$ & $-3.2$ \\
\hline
$1$ & $-2.05$ & $-2.02$ & $-1.3$ \\
\hline
$3$ & $-1.9$ & $-2.4$ & $-0.92$ \\
\hline
$5$ & $-2.0$ & $-2.34$ & $-0.92$ \\
\hline
$7$ & $-1.9$ & $-2.35$ & $-0.89$ \\
\hline
$9$ & $-1.88$ & $-2.335$ & $-1.80$ \\
\hline
$11$ & $-1.85$ & $-2.342$ & $-2.25$ \\
\hline
\end{tabularx}
\end{table}
\subsection{DCAR Performance Evaluation}
\label{sec:dcar}
\begin{figure}[!t]
    \centering
    \subfloat[]{\includegraphics[width=0.83\columnwidth]{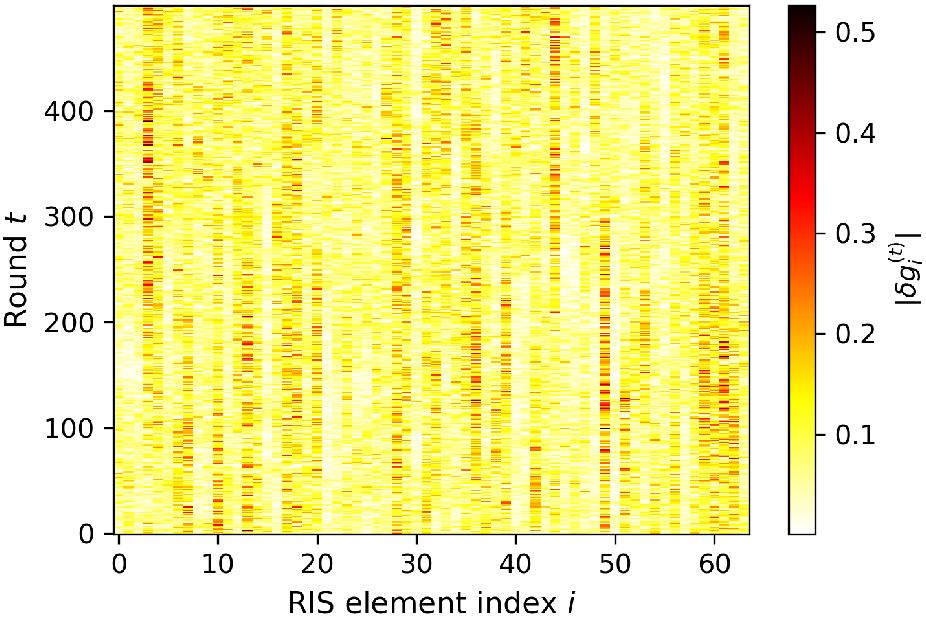}\label{fig:pert1}}
    \\ \vspace{-0.8em}
    \subfloat[]{\includegraphics[width=0.83\columnwidth]{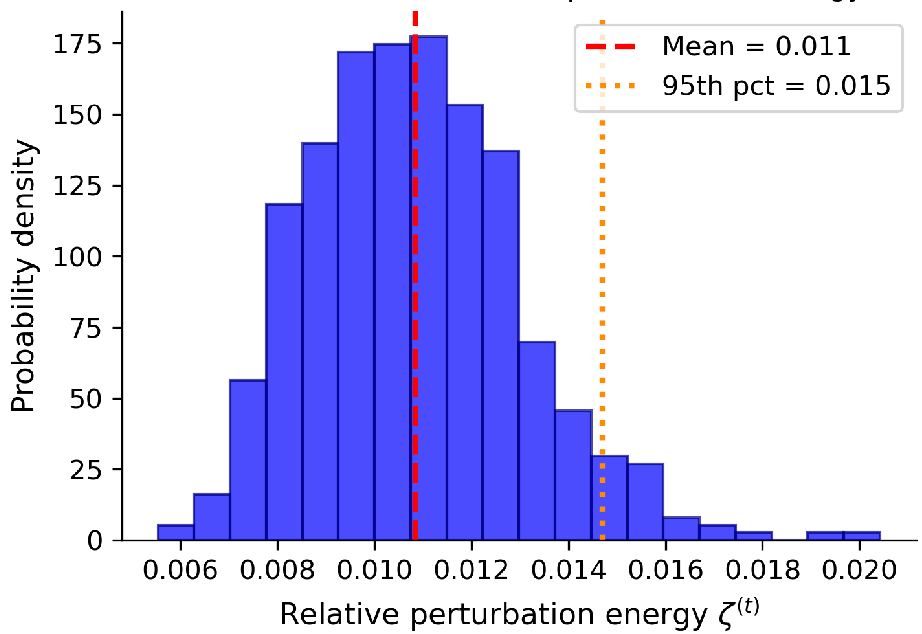}\label{fig:pert2}}
    \caption{Channel perturbation under AR(1) slow fading ($\rho=0.9975$), $N=64$ and $\gamma_0=-20$ dB. (a) Per-element 
    perturbation magnitude $|\delta g_i^{(t)}|$ over $500$ rounds. (b) Distribution of the relative perturbation energy $\zeta^{(t)}=\|\delta_g^{(t)}\|^2/\|g^{(t-1)}\|^2$. The dashed and dotted lines indicate the empirical mean and $95$-th 
    percentile, respectively.}
\label{fig:perturbation_heatmap}
\end{figure}
DCAR reduces pilot overhead by exploiting temporal
correlation, estimating only the perturbation
$\delta_g^{(t)}$ from $K_D\!\ll\!N$ probes rather than
re-estimating all $N$ cascaded channel coefficients each update. The evaluation
has four parts. Sec.~\ref{sec:dcar_heatmap} verifies that
the perturbation energy remains small relative to the channel
energy under slow fading, supporting the differential update
model in~\eqref{eq:dcar_update}. Sec.~\ref{sec:dcar_overhead}
evaluates BF gain and pilot overhead relative to full
retraining for different RIS sizes. Sec.~\ref{subsubsec:kd}
and~\ref{subsubsec:tau_role} examine the effects of the probe
count $K_D$ and threshold $\tau_{\mathrm{DCAR}}$.
Unless stated otherwise, all results use MMSE estimation,
QPSK modulation and slow GM fading
($\rho\!\approx\!0.9975$) with
$\gamma_0\!=\!-20$ dB. The regularization parameters are
$\lambda\approx3.12$ and $\mu\approx0.074$, refer
\eqref{eq:dcar_alpha_lambda}.

\subsubsection{Channel Perturbation Analysis}
\label{sec:dcar_heatmap}
DCAR update model assumes that the channel variation between successive rounds remains small i.e., $|\delta_g^{(t)}|\!\ll\!|g^{(t-1)}|$. Under the AR(1) approximation for $g_i^{(t)}$ (Sec.~\ref{subsec:dcar}), the approximated
perturbation of the $i$-th cascaded-channel coefficient is:
\begin{equation}
\delta g_i^{(t)} = g_i^{(t)} - g_i^{(t-1)} = (\rho-1)g_i^{(t-1)} + \sqrt{1-\rho^2} z_i^{(t)},
\label{eq:delta_gi}
\end{equation}
where $z_i^{(t)}$ is the $i$-th element of the independent innovation vector $Z^{(t)}\!\!\sim\!\!\mathcal{CN}(0,I)$, satisfying 
$E[|z_i^{(t)}|^2]\!\!=\!\!1$. Using $E[|g_i^{(t-1)}|^2]\!\!=\!\!1$ and that $g_i^{(t-1)}\perp\!\!\!\perp z_i^{(t)}$, expected 
per-element perturbation energy is derived as:
\begin{equation}
E[|\delta g_i^{(t)}|^2] = (\rho-1)^2 + (1-\rho^2) = 2(1-\rho).
\label{eq:delta_energy}
\end{equation}
Fig.~\ref{fig:perturbation_heatmap} validates this over $500$ rounds for $N=64$, $\gamma_0=-20$ dB. Fig.~\ref{fig:perturbation_heatmap}(a) shows perturbations
appearing randomly across RIS elements and rounds, with no
persistent spatial or temporal hotspots, consistent with the AR(1)
model's independent innovations. Fig.~\ref{fig:perturbation_heatmap}(b) shows the relative
perturbation ratio $\zeta^{(t)}$ from~\eqref{eq:validation_perturbation}
centered at a mean of $0.011$ (95th-percentile $0.015$, maximum
below $0.021$) -- exceeding the marginal-expectation ratio
$E[|\delta_g^{(t)}|^2]/E[|g^{(t-1)}|^2]=2(1-\rho)=0.005$
from~\eqref{eq:delta_energy}, since $\zeta^{(t)}$ is the ratio of
statistically coupled variables rather than of their mean:
under the AR(1) model~\eqref{eq:GM}, $\delta g_i^{(t)}$ depends
directly on $g_i^{(t-1)}$ via $(\rho\!-\!1)g_i^{(t-1)}$, giving
$E\!\left[\frac{|\delta_g^{(t)}|^2}{|g^{(t-1)}|^2}\right]\!>\!\frac{E[|\delta_g^{(t)}|^2]}{E[|g^{(t-1)}|^2]}$. As $\zeta^{(t)}$
never exceeds $0.021$, perturbation energy stays below $2.1\%$ of
channel energy throughout tracking, supporting the small-perturbation
assumption in~\eqref{eq:dcar_recovery}--\eqref{eq:dcar_update}.
\subsubsection{Overhead Reduction Relative to Full Retraining}
\label{sec:dcar_overhead}
Fig.~\ref{fig:8}(a) compares the BF gain and normalized pilot
overhead of DCAR, for
$N\!\!\in\!\!\{4,16,64,256\}$, against periodic full retraining. The pilot overhead decreases from
$64\%$ at $N\!\!=\!4$ to about $38$--$39\%$ for $N\!\!\ge\!\!64$. For
smaller $N$, the variation metric exceeds $\tau_{{DCAR}}$
more often, triggering more retraining cycles and
higher pilot consumption than for larger $N$.
Further reduction comes from estimating the channel perturbation
from only $K_D$ probe observations instead of full retraining.
The smaller observation set increases perturbation-estimation
error, which accumulates over successive updates and gradually
degrades RIS phase alignment until the next full retraining
cycle. The slight overhead increase from $38\%$ at $N\!=\!64$ to
$39\%$ at $N\!=\!256$ falls within the sampling variability of
$500$ MC realizations and does not indicate a
systematic trend.
\subsubsection{$K_D$ Sensitivity}
\label{subsubsec:kd}
\begin{figure}[!b]
    \centering
    \subfloat[{}]{\includegraphics[width=0.87\columnwidth]{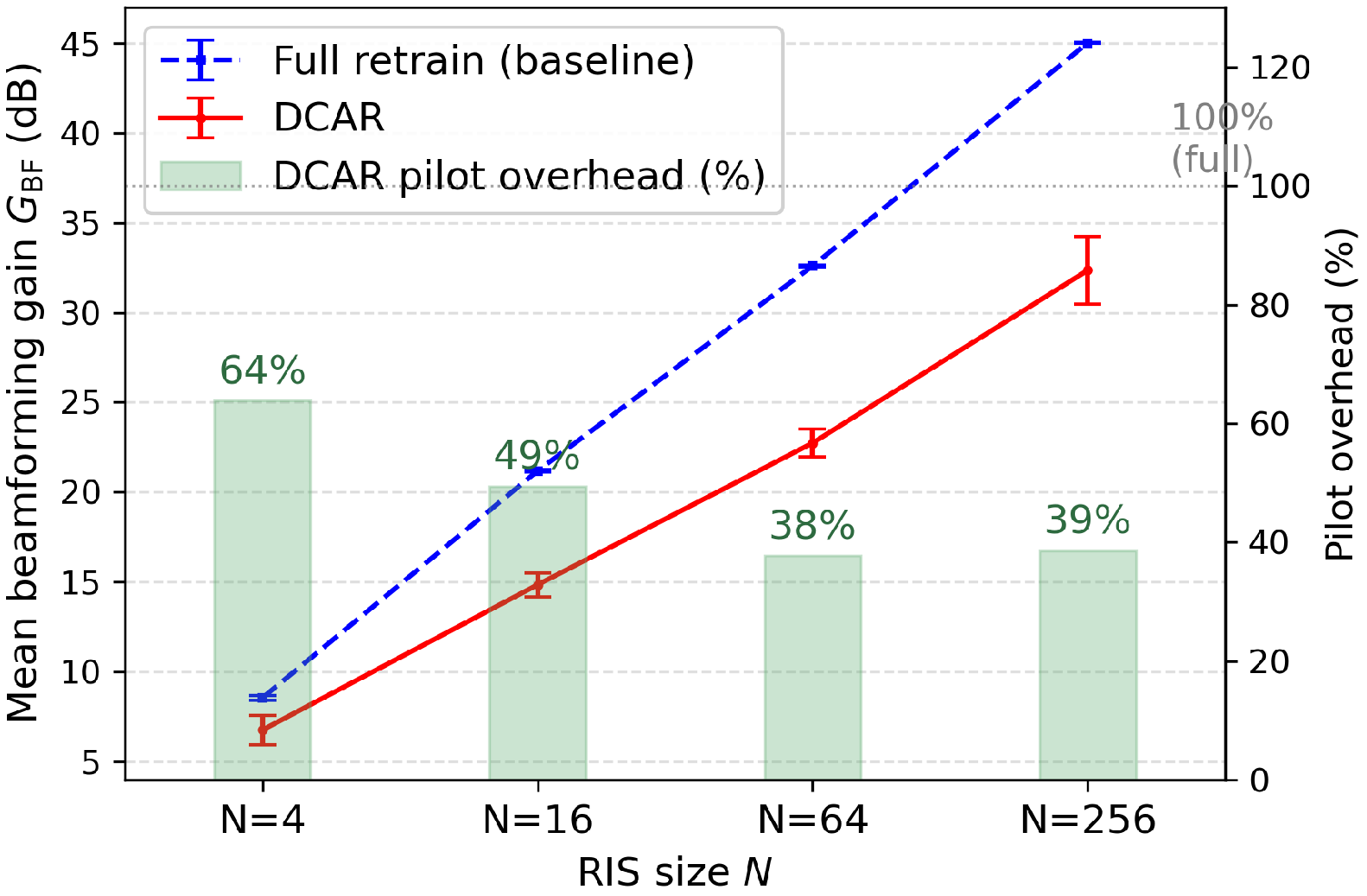}
\label{fig:overhead_reduction}}
\\
\vspace{-.8em}
    \subfloat[{}]{\includegraphics[width=0.87\columnwidth]{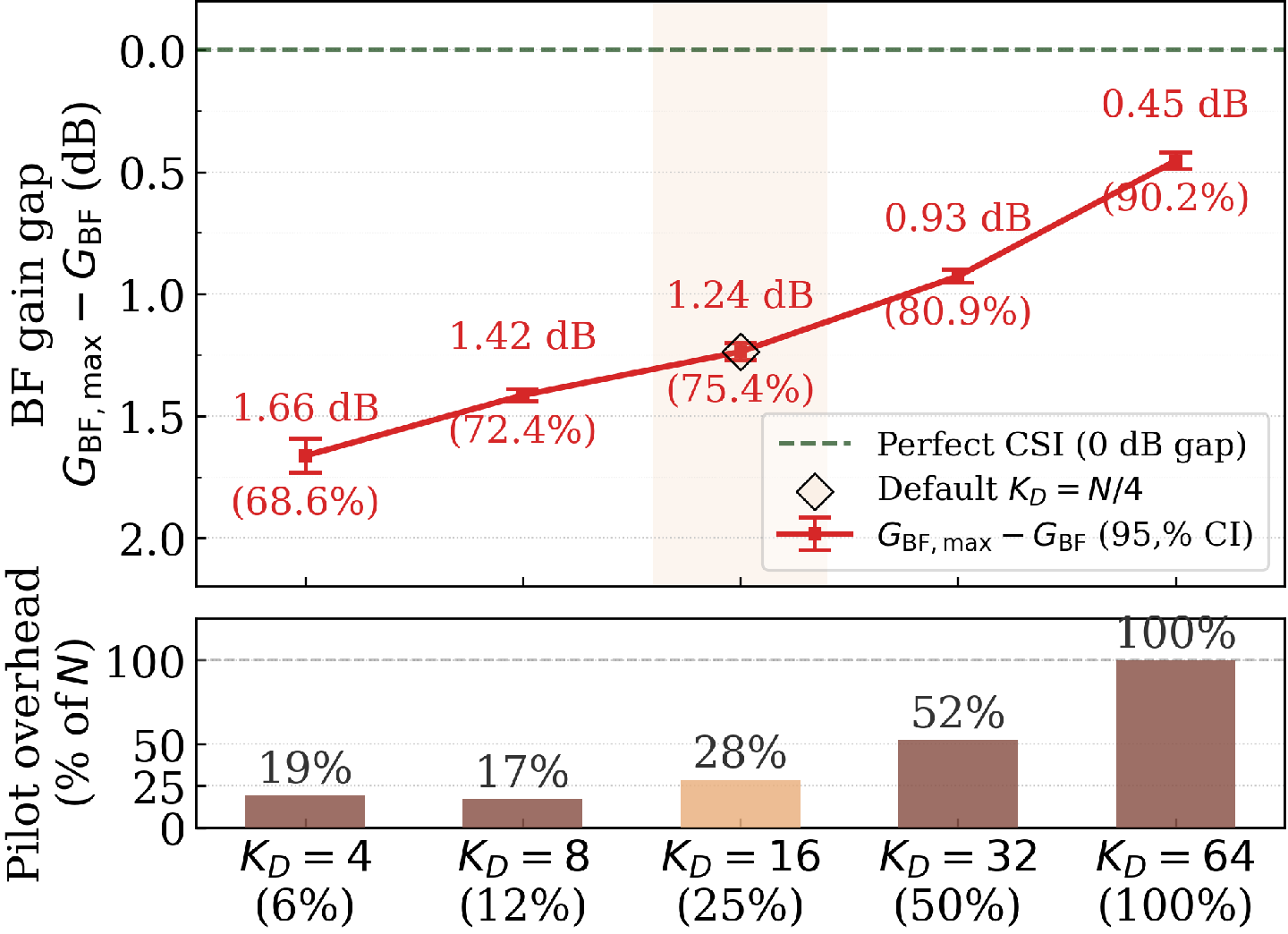}
\label{fig:kd}}
    \caption{(a) Mean BF gain (left axis, 95\% CIs) and normalized pilot overhead (right axis) vs.\ RIS size $N$ under slow AR(1) 
fading ($\rho = 0.9975$). (b)BF gain gap and total pilot overhead as a function of the probe count $K_D$.}
\label{fig:8}
\end{figure}
The probe count $K_D$ trades tracking accuracy against pilot
overhead. By recovering $\delta_g^{(t)}$ relative to an anchor
instead of re-estimating the full channel, DCAR tracks using
fewer than $N$ pilots under slow fading~\cite{refXI,refXIII,refXXIII}.
Fig.~\ref{fig:8}(b) sweeps $K_D\!\in\!\{4,8,16,32,64\}$ for $N\!=\!64$,
$\rho\!\approx\!0.9975$, $\gamma_0\!=\!-20$ dB and $\tau_{{DCAR}}$,
measuring the mean BF gain gap to perfect CSI. Increasing $K_D$
from $4$ to $64$ drops the gap monotonically from $1.66$ dB to
$0.45$ dB, with narrow $95\%$ CIs ($\pm0.033$--$0.069$ dB)
confirming statistical stability. At $K_D\!=\!4$, $v^{(t)}$~\eqref{eq:dcar_metric}
becomes noise-sensitive, triggering $14$ false retrains per $100$
cycles versus $5$ for $K_D\!=\!8$; since each false retrain costs
$N\!=\!64$ extra pilots, this outweighs the probe-count savings. For
$K_D\!\geq\!8$, retraining is governed mainly by $T_{\max}$ and
further increase in $K_D$ yields diminishing returns: raising
$K_D$ from $8$ to $16$ cuts the BF gap by only $0.18$ dB while
overhead rises from $17\%$ to $28\%$ and $K_D\!=\!32$ recovers a
further $0.31$ dB but nearly doubles overhead. The
operating point also depends on $\lambda$ in~\eqref{eq:dcar_alpha_lambda}:
lower SNR or faster variation requires larger $K_D$, higher SNR
allows fewer probes. The default $K_D\!=\!\lfloor N/4\rfloor\!=\!16$
gives a good tradeoff, a $1.24$ dB gap at $28\%$ overhead,
avoiding threshold-driven false retraining.
\subsubsection{Role of $\tau_{{DCAR}}$ under Slow Fading}
\label{subsubsec:tau_role}
The variation threshold $\tau_{{DCAR}}$ safeguards against channel variations that
violate the differential tracking assumption. For $K_D\geq16$,
the EMA-smoothed metric stayed below $\tau_{{DCAR}}=0.75$
throughout, peaking near $v_{{EMA}}\approx0.44$; threshold-triggered
retraining was thus rare and $T_{\max}$ governed retraining
instead. Under this slow-fading regime, $T_{\max}$ sets the
retraining schedule while $\tau_{{DCAR}}$ remains available to
catch larger variations.
\subsection{Comparison with Existing Tracking Methods}
\label{sec:dcar_comparison}
The proposed DCAR algorithm is compared under identical conditions
with two representative RIS channel-tracking baselines that reflect
widely adopted approaches for reducing RIS training overhead: the
Kalman filter (KF)~\cite{refXI} and the ALOCET
framework~\cite{refXXIII}. The three methods represent different
design tradeoffs: Kalman filtering exploits temporal state
estimation at higher computational and memory cost, ALOCET reduces
pilot overhead through adaptive probe selection, while DCAR
estimates channel perturbations using a regularized differential
update to reduce both pilot overhead and computational complexity.

The KF uses Bayesian state estimation, modeling the time-varying channel as a dynamic hidden state observed through noisy pilot measurements rather than a static unknown parameter. It treats the cascaded channel ${g}^{(t)}\!\!\in\!\!\mathcal{C}^{N\!\times\!1}$ as a first-order auto-regressive (AR(1)) process. At interval $t$, the KF maintains an $N\!\!\times\!\!N$ posterior error covariance matrix $\Gamma^{(t|t)}$—which quantifies the uncertainty of the channel estimate—and executes a two-step prediction--correction recursion\footnote{Following standard Kalman filter notation, a quantity indexed as
$(t|t')$ denotes its estimate at interval $t$ using all observations
available up to interval $t'$. Thus, $(t|t-1)$ denotes the predicted
(\emph{a priori}) estimate before the interval-$t$ probe is observed,
whereas $(t|t)$ denotes the corrected (\emph{a posteriori}) estimate
after incorporating the interval-$t$ probe observation.}:
\begin{align}
  \hat{{g}}^{(t|t-1)} &= \rho \hat{g}^{(t-1|t-1)}, \\
  \Gamma^{(t|t-1)} &= \rho^2\Gamma^{(t-1|t-1)} + (1-\rho^2){I}_N, \label{eq:kf_predict}\\
  \kappa^{(t)} &= \Gamma^{(t|t-1)}\Omega^H \!\left(\Omega\Gamma^{(t|t-1)}\Omega^H + \sigma_n^2{I}_{K_D}\right)^{\!-1}, \label{eq:kf_gain}\\
  \hat{{g}}^{(t|t)} &= \hat{{g}}^{(t|t-1)} + \kappa^{(t)}\!\left(\mathbf{y}^{probe,(t)}_{K_D} - \Omega\hat{{g}}^{(t|t-1)}\right), \label{eq:kf_update}
\end{align}
where $\Omega\!=\!\sqrt{\mathcal{P}}\Phi_{K_D}\!\!\in\!\mathcal{C}^{K_D\times N}$ is the effective probe matrix, ${y}^{probe,(t)}_{K_D}\!\!\in\!\mathcal{C}^{K_D\times1}$ is the received probe observation vector and $\kappa^{(t)}\!\!\in\!\mathcal{C}^{N\times K_D}$ is the Kalman gain. The Kalman gain weighs the prediction and probe observations for channel estimate update. Maintaining $\Gamma^{(t|t)}$ requires $\mathcal{O}(N^2)$ memory, so evaluating~\eqref{eq:kf_gain} requires $\mathcal{O}(K_DN^2)$ operations per update interval.

ALOCET does not track state dynamics; instead, it selects a subset
of $K_D$ configurations by minimizing the Cram\'er--Rao lower bound
(CRLB) of the spatial steering vector $\mathbf{m}$, which specifies
the minimum estimation variance achievable by any unbiased estimator
for a given set of observations. With
$\mathbf{C}_{\mathrm{FIM}}(\hat{\mathbf{m}},\psi_i)\!\in\!\mathbb{R}^{2\times2}$
denoting the individual Fisher information matrix (FIM) block
contributed by the $i$-th configuration candidate $\psi_i$, it
chooses indices $\mathcal{S}^*$ minimizing this bound as
\begin{equation}
  \mathcal{S}^* = \operatornamewithlimits{arg\min}_{|\mathcal{S}|=K_D} \operatorname{Tr}\!\left( \sum_{i\in\mathcal{S}}\mathbf{C}_{\mathrm{FIM}}(\hat{\mathbf{m}},{\psi}_i) \right)^{\!-1},
  \label{eq:alocet_crlb}
\end{equation}
ALOCET tracks the channel geometrically based on a two-dimensional steering vector $\mathbf{m}\!\!=\!\![m_x, m_y]^T$, this block is structurally restricted to a $2\!\!\times\!\!2$ matrix corresponding to the spatial azimuth and elevation coordinates\footnote{The original ALOCET method is derived for geometry-based channel models. Since the Rayleigh channel considered here does not provide the required geometric information, the evaluated baseline uses an energy-based approximation for adaptive probe selection.}. The functions $q(\cdot)$ the directional sensitivity metrics,
quantify the variation of the received signal power related
to small angular displacements along these spatial coordinates \cite{refXXIII}:
\begin{equation}
  \mathbf{C}_{\mathrm{FIM}}(\hat{\mathbf{m}},{\psi}_i) = 16\pi^2\varepsilon
  \begin{bmatrix}
    q_{\bar{\Psi}_{xx}}(\iota(\tilde{\mathbf{m}})) & q_{\bar{\Psi}_{xy}}(\iota(\tilde{\mathbf{m}})) \\
    q_{\bar{\Psi}_{yx}}(\iota(\tilde{\mathbf{m}})) & q_{\bar{\Psi}_{yy}}(\iota(\tilde{\mathbf{m}}))
  \end{bmatrix}.
  \label{eq:alocet_fim_block}
\end{equation}

Here, $\varepsilon$ is a constant scaling factor and
$q_{\bar{\Psi}_{ij}}(\cdot)$ denotes the second-order directional
sensitivity terms used in the ALOCET formulation~\cite{refXXIII}. The
mapping $\iota(\cdot)$ transforms the angular displacement vector
$\tilde{\mathbf{m}}$ into the corresponding steering-response
coordinate. Evaluating~\eqref{eq:alocet_crlb} over all $N$ candidate
configurations requires $\mathcal{O}(N^2)$ operations per round.

\begin{figure}[!t]
\centering
\subfloat[]{%
\includegraphics[width=.77\columnwidth]{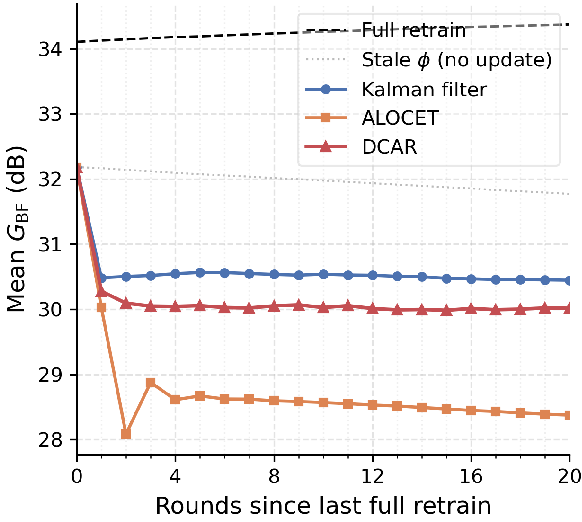}
\label{fig:dcar_rounds}}
\\
\vspace{-1em}
\subfloat[]{%
\includegraphics[width=.77\columnwidth]{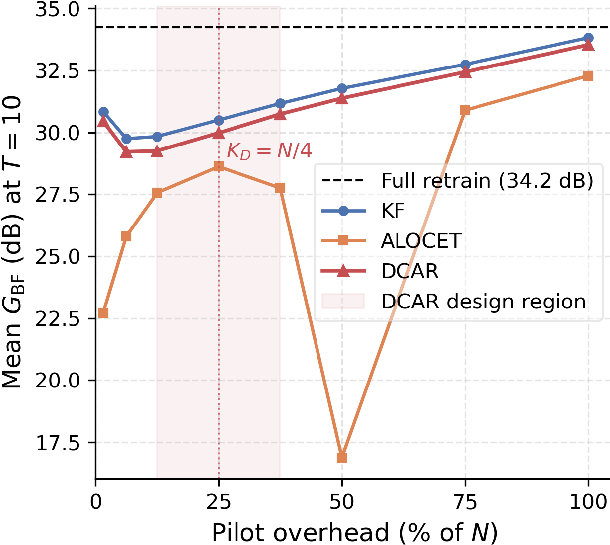}
\label{fig:dcar_overhead}}
\\
\vspace{-1em}
\subfloat[]{%
\includegraphics[width=.78\columnwidth]{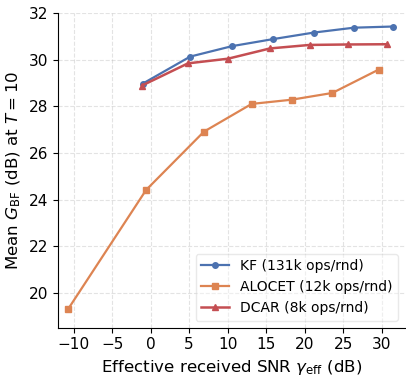}
\vspace{-1em}
\label{fig:dcar_complexity}}
\caption{Performance comparison of DCAR, Kalman filter and ALOCET:
(a)~Mean BF gain $G_{\mathrm{BF}}$ over tracking rounds;
(b)~$G_{\mathrm{BF}}$ versus pilot overhead;
(c)~$G_{\mathrm{BF}}$ versus effective received SNR and reported computational cost. \textit{Note:} All methods are evaluated for identical conditions: $N\!\!=\!\!64$-element RIS, $\rho\!\!=\!\!0.9975$, $\gamma_0\!\!=\!\!-20$ dB, pilot budget $K_D\!\!=\!\!16$ and $M\!\!=\!\!500$ MC realizations.}
\label{fig:dcar_comparison}
\end{figure}

DCAR avoids full retraining and combinatorial row se-
lection, applying scalar MMSE shrinkage factors $(\lambda,\mu)$ to
reduce the effective noise amplification from $N/K_D$ to $\mu
N/K_D$. It requires only $\mathcal{O}(NK_D)$ operations
and $\mathcal{O}(N)$ memory per update interval, with $\rho$
used only during one-time offline initialization. Table~\ref{tab:tracking_comparison} and Fig.~\ref{fig:dcar_comparison}
summarize the performance and design tradeoffs. 
In
Fig.~\ref{fig:dcar_comparison}(a), all methods
start with full retrain at round T=0 [11], [12]. By $T\!=\!20,$
the KF reaches 30.4 dB via optimal AR(1) prediction. DCAR
achieves 30.0 dB by suppressing noise via scalar shrinkage,
keeping within 0.4 dB of the KF. ALOCET does not exploit
temporal correlation or differential channel updates, so only
gains upto 28.3 dB.

Fig.~\ref{fig:dcar_comparison}(b) compares performance across
different pilot budgets $K_D$. At the default $25\%$ overhead,
DCAR achieves $30.0$ dB, close to the Kalman filter ($30.4$ dB).
ALOCET exhibits a localized degradation near
$K_D\!\approx\!N/2$ before recovering to $32.3$ dB as the pilot
budget approaches full retraining. Although ALOCET evaluates the full RIS aperture through
an $\mathcal{O}(N^2)$ scoring stage, DCAR consistently achieves
higher beamforming gain through regularized differential updates\footnote{In ALOCET, only the
selected pilot coefficients are refreshed during each tracking
round, while the remaining coefficients retain their estimates
from the previous full retraining. When
$K_D\!\approx\!N/2$, the updated and retained coefficients
contribute nearly equal power to the beamforming vector. As the
retained coefficients gradually drift from the true channel,
their phase mismatch with the updated coefficients reduces the
overall coherent combining gain, producing the localized dip.
For smaller or larger values of $K_D$, one group dominates the
combined response, making this effect much less pronounced. This
behavior occurs outside the default operating point
$K_D=\lfloor N/4\rfloor=16$ used throughout the remainder of this
paper.}.
\begin{table}[!t]
\centering
\caption{Per-round complexity and performance summary of the compared
tracking methods.}
\label{tab:tracking_comparison}
\renewcommand{\arraystretch}{1.1}
\setlength{\tabcolsep}{2pt}
\begin{tabularx}{\columnwidth}{|X|c|c|c|c|}
\hline
\textbf{Method} &
\textbf{BF Gain} &
\textbf{QPSK BER}$^\dagger$ &
\textbf{Operations/Round} &
\textbf{Memory} \\
\hline
Perfect retrain
& $34.4$ dB & $7.7\!\times\!10^{-8}$
& $\mathcal{O}(N^2)+N$ pilots & $\mathcal{O}(N)$ \\
\hline
KF~\cite{refXI}
& $30.4$ dB & $4.6\!\times\!10^{-4}$
& $\mathcal{O}(K_DN^2)$ & $\mathcal{O}(N^2)$ \\
\hline
ALOCET~\cite{refXXIII}
& $28.3$ dB & $4.7\!\times\!10^{-3}$
& $\mathcal{O}(N^2)+\mathcal{O}(NK_D)$ & $\mathcal{O}(N)$ \\
\hline
\textbf{DCAR}
& $30.0$ dB & $7.8\!\times\!10^{-4}$
& $\mathcal{O}(NK_D)$ & $\mathcal{O}(N)$ \\
\hline
\end{tabularx}
{\footnotesize~\raggedright $^\dagger$QPSK BER is evaluated as
$P_b = \frac{1}{2}\operatorname{erfc}\!\bigl(\sqrt{E_b/N_0}\bigr)$~\cite{refXXIV},
where $E_b/N_0 = \gamma_{\mathrm{eff}}/2$ for QPSK (two bits per symbol) and
$\gamma_{\mathrm{eff}}$~\eqref{eq:effective_snr} uses the BF gain
achieved by each method at $T=20$ rounds.\par}
\end{table}

Fig.~\ref{fig:dcar_comparison}(c) compares BF gain versus
effective received SNR. At low SNR, KF and DCAR improve
faster than ALOCET, whose probe selection is less reliable
under noisy estimates. The gap narrows with increasing SNR,
DCAR staying near the KF at only $\mathcal{O}(NK_D)$
operations versus $\mathcal{O}(K_DN^2)$ for the Kalman filter
(Table~\ref{tab:tracking_comparison}).

\subsection{Discussion}
\label{discuss}
Unlike SimRIS~\cite{refXVII} and Sionna~\cite{refXVIII}, which
execute propagation and communication models within a single
simulation environment and primarily report propagation- or
link-level performance metrics (Table~\ref{tab:sota_comparison}),
TRACE realizes the transmitter, channel, controller and receiver
as independent processes connected through UDP-based control-
and data-plane interfaces, additionally recording controller-level
statistics, channel estimates, beamforming gain, NMSE,
pilot-overhead statistics, RIS reconfiguration history and
timestamped experiment logs for offline analysis of closed-loop
RIS adaptation. Using the default DCAR configuration
($K_D=\lfloor N/4\rfloor$), the average control-plane round-trip
time over $500$ tracking intervals is $5.2$ ms ($\approx0.32$ ms
per probe), while a full $N=64$ retraining cycle requires $20.0$
ms (well within the coherence intervals of Sec.~\ref{sec:setup}),
showing the socket-based architecture adds only modest software
overhead while supporting closed-loop operation. These timings
do not affect the reported pilot-overhead reductions, which
depend only on pilot count; execution time here is dominated by
the configured inter-probe pacing interval rather than software
processing. The evaluation assumes an ideal continuous phase-only
RIS, single-antenna Tx--Rx links and simulated channel evolution
without an RF front end, providing a reproducible environment for
evaluating CE, tracking and communication algorithms independent
of specific hardware implementations\footnote{The TRACE framework and DCAR
implementation will be released as open-source software upon
acceptance to support reproducible RIS research.}.

\section{Conclusion}
\label{sec:conclusion}
This paper addresses two challenges in RIS-assisted communication:
the lack of a common experimental framework for evaluating channel
estimation, tracking and communication strategies under identical
conditions and the pilot overhead incurred by repeated RIS
retraining under slowly-varying channels. To address these
challenges, we presented TRACE, a modular socket-based framework
that decouples the transmitter, radio environment, controller and
receiver through independent control- and data-plane interfaces,
enabling interchangeable channel-estimation, channel-evolution,
tracking and communication modules to be evaluated within a common
experimental pipeline. Unlike algorithm-specific implementations, TRACE supports
integration of alternative channel-estimation, tracking and
communication modules by replacing only the corresponding
processing module while preserving the existing socket
interfaces and message exchange. The current implementation
demonstrates this through interchangeable MMSE and OMP
channel-estimation and the DCAR tracking module. Within
TRACE, the proposed differential channel-aware RIS update (DCAR)
algorithm exploits temporal correlation to estimate channel
perturbations from reduced probe observations through a
regularized differential update. The framework was demonstrated
through configuration-level module substitution using minimum mean
square error and orthogonal matching pursuit channel-estimation,
Gauss--Markov and Gaussian random walk channel evolution, BPSK,
QPSK and 16-QAM modulation and multiple RIS sizes. Across the
evaluated scenarios, DCAR reduced pilot overhead while maintaining
beamforming performance close to a Kalman-filter-based tracker at
significantly lower computational complexity. The current
implementation realizes all modules as software processes
communicating through UDP sockets on a single host, providing a
reproducible platform for algorithm development and evaluation.
Owing to TRACE's modular architecture, it provides a foundation for future extensions to SDR platforms, programmable RIS hardware, adaptive pilot allocation and larger-scale multi-RIS or MIMO deployments.

\end{document}